\documentclass[aps,english,preprint,nofootinbib,preprintnumbers,superscriptaddress]{revtex4-2}
\usepackage{mathrsfs}
\usepackage{hyperref}
\hypersetup{
    colorlinks=true,
    linkcolor=blue,
    filecolor=magenta,      
    urlcolor=cyan,
    pdftitle={Overleaf Example},
    pdfpagemode=FullScreen,
    }
\usepackage{graphicx}
\usepackage{amsmath, amssymb}
\usepackage{babel}
\usepackage{color}
\usepackage{empheq}
\usepackage{slashed}
\usepackage[T1]{fontenc}
\usepackage{titlesec}
\def\beqn{\begin{eqnarray}}
\def\eeqn{\end{eqnarray}}
\def\barr{\begin{array}}
\def\earr{\end{array}}
\def\btab{\begin{tabular}}
\def\etab{\end{tabular}}
\def\bite{\begin{itemize}}
\def\eite{\end{itemize}}
\def\bcen{\begin{center}}
\def\ecen{\end{center}}

\def\ba#1\ea{\begin{align}#1\end{align}}

\begin{document}

\title{A comprehensive theory framework for perturbative calculations of $\delta_C$ in superallowed beta decays}

\author{Chien-Yeah Seng}

\affiliation{Department of Physics and Astronomy, University of Tennessee, Knoxville, TN 37996, USA}

\date{\today}

\begin{abstract}

The present high precision determination of $|V_{ud}|$ from superallowed beta decays of $J^P=0^+$, $T=1$ nuclei is largely built on top of the ``standard'' calculation of the isospin breaking correction $\delta_C$ to the Fermi matrix element, which assumes the splitting $\delta_C=\delta_{C1}+\delta_{C2}$, where $\delta_{C1}$ and $\delta_{C2}$ represent the ``isospin mixing'' correction and the ``radial mismatch'' correction, respectively. In this paper I show that this  formalism violates the rule of nucleon basis independence, similar to the gauge invariance requirement in quantum field theory, therefore its outcome is inevitably subject to uncontrolled systematic errors. I further argue that the perturbative formalism is the only approach that allows the computation of $\delta_C$ at the precision level required for the test of the Cabibbo unitarity. Advancing from existing literature, I develop the full ``generating function approach'' to compute $\delta_C$ perturbatively, with connections to nuclear mass splittings and the isospin breaking in nuclear charge radii as theory benchmarks.

\end{abstract}

\maketitle
\newpage


\section{Introduction}

In 1963, Cabibbo postulated that the strangeness-conserving and strangeness-changing charged weak currents are mixed through a $2\times 2$ matrix characterized by a single rotation angle $\theta_C$, the Cabibbo angle~\cite{Cabibbo:1963yz}. In the Standard Model (SM), the theorem above translates into the statement that up to $10^{-4}$ precision (where $|V_{ub}|^2$ is negligible~\cite{ParticleDataGroup:2024cfk}), the absolute values of the two top-row Cabibbo-Kobayashi-Maskawa (CKM)~\cite{Kobayashi:1973fv} matrix elements are expressed entirely in terms of the  Cabibbo angle, $|V_{ud}|=\cos\theta_C$, $|V_{us}|=\sin\theta_C$, and thus satisfy the relation $|V_{ud}|^2+|V_{us}|^2=1$ known as the Cabibbo unitarity. It represents one of the SM predictions that can be tested to extremely high precision at low energies, which imposes strong constraints on physics beyond the Standard Model (BSM), for example the non $V-A$ charged weak interactions between quarks and leptons, at multi-TeV level~\cite{Cirigliano:2012ab}.

Given the absolute size $|V_{ud}|^2\approx 0.95$, in order to test the Cabibbo unitarity at permille level, one requires the same level of precision in the determination of $|V_{ud}|^2$, an order of magnitude more than $|V_{us}|^2$. At the present, the primary channels for the extraction of $|V_{ud}|$ include the semileptonic pion decay, free neutron decay and nuclear beta decays. The semileptonic pion decay is the theoretically cleanest channel, thanks to recent lattice calculations of structure-dependent radiative corrections~\cite{Feng:2020zdc,Yoo:2023gln} and improved calculation of higher-order quantum electrodynamics (QED) effects~\cite{Cirigliano:2026ios}; however, it is currently limited by the uncertainty in the branching ratio~\cite{Pocanic:2003pf}, which will be improved in future experiments~\cite{PIONEER:2022alm,PIONEER:2022yag}. Free neutron decay has similarly benefited from the recent progress in structure-dependent radiative corrections~\cite{Seng:2018yzq,Seng:2018qru,Seng:2020wjq,Czarnecki:2019mwq,Shiells:2020fqp,Hayen:2020cxh,Ma:2023kfr} and higher-order QED effects~\cite{Cirigliano:2023fnz,VanderGriend:2025mdc,Cao:2025lrw}; however, experimental discrepancies in the neutron lifetime and the axial coupling constant~\cite{ParticleDataGroup:2024cfk} limit the precision of its $|V_{ud}|$ determination. Nuclear beta decays, on the other hand, possess the advantage of having multiple transitions to be averaged over, which substantially reduces the experimental uncertainty. As a price to pay, they are subject to complicated nuclear structure-dependent effects that contribute to the theory uncertainty in the extracted $V_{ud}$.

Recent years have seen a $\sim 3\:\sigma$ deficit in the Cabibbo unitarity, which provides hints for BSM physics, and also motivates the community to scrutinize the present accepted values of $|V_{ud}|$~\cite{Brodeur:2023eul}. Over the past five decades, the heroic effort by Hardy and Towner in compiling countless experiment results and carefully analyzing all theoretical inputs in superallowed $0^+\rightarrow 0^+$ beta decays of $T=1$ nuclei, has established this decay channel as the best avenue for the precise extraction of $|V_{ud}|$ ~\cite{Hardy:1975eq,Hardy:1990sz,Hardy:2004id,Hardy:2008gy,Hardy:2014qxa,Hardy:2020qwl}. The robustness of such  extraction relies critically on the quality of all the SM theory inputs, especially the nuclear-structure-dependent corrections, that are applied to the decays. They enter (1) the Fermi function and shape factor, (2) the radiative corrections, and (3) the isospin symmetry breaking (ISB) corrections. For (1), based on a model-independent relation between the nuclear charge and weak form factors~\cite{Seng:2022inj}, the combination of experimental charge radii and \textit{ab initio} calculation of moment ratios are able to pin down the nuclear size and shape effects~\cite{Seng:2023cgl,He:2026zie}. For (2), recent developments of the current algebra~\cite{Seng:2022cnq,Gorchtein:2023naa} and effective field theory approach~\cite{Cirigliano:2024msg} open the window for reliable \textit{ab initio} calculations of nuclear-structure-dependent radiative corrections, which were successfully applied to light superallowed transitions~\cite{Gennari:2024sbn,Cirigliano:2024rfk,King:2025fph} and can also be extended to heavier ones. This leaves (3), the ISB correction as the final object to scrutinize, and is the focus of this paper.

The ISB correction $\delta_C$ is defined through the deviation of the Fermi matrix element $M_F$ from its isospin-symmetric limit:
\begin{equation}|M_F|^2=|M_F^0|^2(1-\delta_C)\ ,\label{eq:deltaCdef}
\end{equation}
where $|M_F^0|^2=2$ for superallowed beta decays  of $T=1$ nuclei. Theory calculations of $\delta_C$ have spanned almost 7 decades~\cite{MacDonald:1958zz}, examples include calculations based on the nuclear shell model with Woods-Saxon potential~\cite{Towner:2002rg,Towner:2007np,Hardy:2008gy,Xayavong:2017kim,Xayavong:2025jdh}, Hartree-Fock wave functions~\cite{Ormand:1989hm,Ormand:1995df}, density functional theory~\cite{Satula:2011br,Satula:2016hbs}, random-phase approximation~\cite{Liang:2009pf} and the isovector monopole resonance sum rule~\cite{Auerbach:2008ut}. These different calculations in general give very different predictions for $\delta_C$, some even differ by orders of magnitude.
Due to the high impact of the $0^+\rightarrow 0^+$ critical reviews~\cite{Hardy:1975eq,Hardy:1990sz,Hardy:2004id,Hardy:2008gy,Hardy:2014qxa,Hardy:2020qwl}, it is almost always the case that, when the value of $|V_{ud}|$ from superallowed beta decays is quoted, the implicit theory inputs for $\delta_C$ has been the shell-model calculation with Woods-Saxon potential. Calculations of this type are built upon the following fundamental assumption, which I thereby denote as the ``C1C2 formalism''. In this formalism, the ISB correction splits into two pieces: $\delta_C=\delta_{C1}+\delta_{C2}$, where $\delta_{C1}$ represents the ``isospin mixing'' correction of the external nuclear states, whereas $\delta_{C2}$ is the ``radial mismatch'' correction originates from the slight mismatch between the proton and neutron radial wavefunction (with the same radial quantum number); numerically, one always finds $\delta_{C2}>\delta_{C1}$. A strong argument favoring the C1C2 + Woods-Saxon calculation is that it is the only approach that successfully aligns the $\mathcal{F}t$-values from different superallowed decays, a consequence of the conserved vector current (CVC) hypothesis in the SM~\cite{Towner:2010bx}. The quoted uncertainties in $\delta_{C}$ from such calculations are quite small, and as a result, the dominant quoted theory uncertainty of $|V_{ud}|$ given in the critical reviews above come from the nuclear-structure-dependent radiative corrections instead of $\delta_C$.

For many years, there has been speculation about whether the quoted uncertainties in the C1C2 + Woods-Saxon result of $\delta_C$ are robust. In particular, Miller and Schwenk~\cite{Miller:2008my,Miller:2009cg} pointed out that the separation $\delta_C=\delta_{C1}+\delta_{C2}$ is model-dependent and violates the basic commutation relations of isospin ladder operators, and restoring the commutation relation could lead to a substantial reduction of the size of $\delta_C$. 
In this paper, I will rephrase the Miller-Schwenk argument in a more transparent way, by showing that the dominant contribution to $\delta_C$ in the C1C2 formalism, namely the radial mismatch correction, is constructed on an arbitrary choice of the single proton and neutron basis wavefunction, which physical observables cannot depend on. The removal of this basis dependence requires an additional term which is missing in the C1C2 formalism, as pointed out in Ref.\cite{Miller:2009cg}. Subsequently, the quoted averaged value of $|V_{ud}|$ from superallowed decays based on the $\delta_C$ inputs from such a method cannot avoid from large systematic errors. 

In recent years, nuclear \textit{ab initio} methods have emerged as more reliable approach to compute $\delta_C$ from first principles. The working principle is simply to include the most general ISB potential in the nuclear Hamiltonian, and use it to compute the full Fermi matrix element $M_F$ in order to extract $\delta_C$ from its definition $|M_F|^2=2(1-\delta_C)$. This approach automatically accounts for all contributions to $\delta_C$ and is in principle rigorous. Attempts along this direction includes the use of No-Core Shell Model (NCSM)~\cite{Caurier:2002hb} and Quantum Monte Carlo (QMC)~\cite{Piarulli:2026abq} techniques for the lightest superallowed transition, ${}^{10}\text{C}\rightarrow{}^{10}\text{B}$.
The critical shortcoming of this approach, however, is that $M_F\sim \mathcal{O}(1)$, so one requires a $10^{-4}$ accuracy level in nuclear many-body calculations for the outcome to be useful, which is practically very challenging. Stochastic methods such as QMC or nuclear lattice effective field theory (NLEFT)~\cite{Lahde:2019npb} are limited by statistical uncertainties, whereas for deterministic methods such as NCSM, coupled-cluster (CC) or in-medium similarity renormalization group (IMSRG), approximations made in the many-body methods often introduce spurious ISB effects which are difficult to separate from the physical $\delta_C$~\cite{Farren:2024spl}. 

A way to circumvent the difficulty above is to avoid computing the full Fermi matrix element $M_F$, but instead to tackle directly the leading expression that contributes to the physical $\delta_C$, which can be derived from a perturbation theory. One then proceed to compute the nuclear matrix element associated to this leading expression. The initial framework of this approach was developed in Refs.\cite{Seng:2022epj,Seng:2023cvt}. Certain simplifications were made in those papers, for example the isovector dominance of the ISB potential was assumed, and Coulomb potential took the form of a uniformly charged sphere. Such approximations allowed a straightforward connection between $\delta_C$ and experimental observables such as nuclear charge radii. In this work, I develop further on the existing result and present a fully generalized theory framework for perturbative calculations of $\delta_C$ that removes all unnecessary approximations. In this framework, the object of interest are ``generating functions'' that consist of ground-state nuclear matrix elements of a nuclear Green's function sandwiched by two ISB potentials. The ISB potential can be obtained through commutation relations between isospin ladder operators and the full Hamiltonian. 

Finally, benchmarking with measurable quantities is a necessary procedure to ensure the accuracy of theory predictions. In this paper I propose a two-step benchmarking process: (1) Benchmark the ISB potential $V$ with the isobaric multiplet mass equation (IMME) that describes the ISB-induced nuclear mass differences~\cite{Lam:2013bhc}, and (2) Benchmark the nuclear Green's function $1/(E-H)$ with the ISB effect in nuclear charge radii~\cite{Seng:2022epj}. The latter, in particular, builds a natural connection between the precision tests of the SM and experimental studies of nuclear size through muonic atom and laser spectroscopy. 

\section{Fermi transition operator in the SM}

To motivate the following discussions, I start by revisiting the basic mathematical formulation of the beta decay matrix element in the SM using a Quantum Field Theory (QFT) setup, and show how it reduces to the Fermi matrix element in quantum mechanics (QM) that involves the isospin ladder operators. 
Throughout this paper, I adopt the ``nuclear physics'' convention of isospin, where $T_3(n)=+1/2$, which is opposite to the particle physics convention.

\subsection{Isospin operators\label{sec:isospin}}

In the SM, one groups the down and up quark field operators as an ``isospin doublet''
\begin{equation}
	q(x)=(d(x)\ ,\ u(x))^{T}\ ,
\end{equation}
which transforms under the isospin rotation as:
\begin{equation}
q(x)\rightarrow Uq(x)\ ,
\end{equation}
where $U$ is an SU(2) matrix. One can define the ``isospin current'' as: 
\begin{equation}
	J^{\mu}_{i}(x)\equiv\bar{q}(x)\frac{\tau_{i}}{2}\gamma^{\mu}q(x)\ ,
\end{equation}
where $\tau_{i}$ is the Pauli matrix in the $d-u$ isospin space. The generators of the SU(2) isospin rotation (or the ``isospin charges'') are just the spatial integral of the time component of these currents:
\begin{equation}
	T_{i}\equiv\int d^{3}xJ^{0}_{i}(0,\vec{x})=\int d^{3}xq^{\dagger}(\vec{x})\frac{\tau_{i}}{2}q(\vec{x})\ ,
\end{equation}
they satisfy the following commutation relation:
\begin{equation}
	[T_{i},T_{j}]=i\varepsilon_{ijk}T_{k}\ ,
\end{equation}
which is just the standard SU(2) Lie algebra. Note that
this is an operator relation that remains true regardless of whether the actual physical system conserves isospin. Finally, the isospin raising and lowering operators are defined as:
\begin{equation}
	T_{\pm}\equiv T_{1}\pm iT_{2}=\int d^3 xq^\dagger(\vec{x})\tau_\pm q(\vec{x})\ ,
\end{equation}
where $\tau_\pm=(\tau_1\pm i\tau_2)/2$. They satisfy the following commutation relation:
\begin{equation}
	[T_{+},T_{-}]=2T_{3}\ .
\end{equation}

Now, I will use $H$ to denote the full Hamiltonian that describes both the strong and electromagnetic
interactions. It commutes with $T_{3}$:
\begin{equation}
	[H,T_{3}]=0\ ,
\end{equation}
which means the total number of down and up quarks are separately conserved; however, it does not commute with $T_\pm$, which means the down and up quarks are not interchangeable, so the full isospin symmetry is not a good symmetry  of $H$. Nevertheless, one can split $H$ into:
\begin{equation}
	H\equiv H_{0}+V\ ,
\end{equation}
where $H_{0}$ is the ``isospin-symmetric'' piece which commutes with all the three isospin operators:
\begin{equation}
	[H_{0},\vec{T}]=0\ ,
\end{equation}
and $V$ is the piece that breaks isospin symmetry.

I will also take this opportunity to introduce some notations for nuclear states that will be adopted later. In general, I use $|n\rangle$, $|n)$ to denote the energy eigenstates of $H$
and $H_{0}$, with energy eigenvalues $E_{n}$ and $E^{0}_{n}$ respectively:
\begin{equation}
	H|n\rangle=E_{n}|n\rangle\ ,\ H_{0}|n)=E^{0}_{n}|n)\ .
\end{equation}
Since the eigenstates of the full Hamiltonian $H$ have definite $T_{3}$
but no definite $T$, I will label the group of three $J^P=0^+$ nuclear ground states that
participate in the superallowed transition as $|g;T_{3}\rangle$, where $g$
is the group label, and $T_{3}=\pm1,0$. Their energies are denoted
as $E_{g;T_{3}}$. For example, for $A=26$ system, one has\footnote{The phase convention of states in this paper follows $T_+|T,T_{3}\rangle=\sqrt{(T-T_3)(T+T_3+1)}|T,T_3+1\rangle$. A different choice of phase will lead to an extra phase factor in the Fermi matrix element and other transition matrix elements.}:
\begin{equation}
	|^{26}\text{Si}\rangle=|g;-1\rangle\ ,\ |^{26m}\text{Al}\rangle=|g;0\rangle\ ,\ |^{26}\text{Mg}\rangle=|g;+1\rangle\ .
\end{equation}
In the isospin-symmetric limit (i.e. $V\rightarrow 0$), these states have exact isospin $T=1$ and their energies become degenerate:  $|g;T_{3}\rangle\rightarrow|g;1,T_{3})$
and $E_{g;T_{3}}\rightarrow E^{0}_{g}$.

\subsection{QFT vs QM matrix elements}

To describe the matrix element of a beta decay process, one needs to first specify the external states. In QFT, the external state of a particle $\phi$ is expressed as plane wave states that are normalized relativistically:
\begin{equation}
	_{\text{QFT}}\langle\phi(\vec{p}')|\phi(\vec{p})\rangle_{\text{QFT}}=(2\pi)^{3}2E_{p}\delta^{3}(\vec{p}-\vec{p}')\ ,
\end{equation}
with $E_{p}=\sqrt{M^{2}_{\phi}+\vec{p}^{2}}$. The plane wave states are eigenstates of the momentum operator $\hat{\vec{p}}$, such that:
\begin{equation}
	\hat{\vec{p}}|\phi(\vec{p})\rangle_{\text{QFT}}=\vec{p}|\phi(\vec{p})\rangle_{\text{QFT}}\ .
\end{equation}
However, in actual nuclear theory calculations, instead of plane wave states one deals with the QM external state $|\phi_{\vec{p}}\rangle$, which represents a wave packet with the center of mass momentum $\vec{p}$, and is normalized as:
\begin{equation}
	\langle\phi_{\vec{p}}|\phi_{\vec{p}}\rangle=1\ .
\end{equation}
It can be constructed from the plane wave state as:
\begin{equation}
	|\phi_{\vec{p}}\rangle=\frac{1}{\sqrt{2E_p}}\int\frac{d^{3}k}{(2\pi)^{3}}\ f(\vec{k}-\vec{p})|\phi(\vec{k})\rangle_{\text{QFT}}\ ,
\end{equation}
where $f(\vec{k})$ is a wave packet function. Assuming that  $|f(\vec{k})|^{2}$
is sufficiently peaked at $\vec{k}=\vec{0}$, such that one can replace: 
\begin{equation}
|f(\vec{k}-\vec{p})|^2 A(\vec{k})\approx |f(\vec{k}-\vec{p})|^2 A(\vec{p})\ ,\label{eq:WPpeak}
\end{equation}
 I obtain its normalization condition as:
\begin{equation}
	 \int\frac{d^{3}k}{(2\pi)^{3}}\ |f(\vec{k})|^{2}=1\ .\label{eq:WPnormalize}
\end{equation}
It is important to note that $|\phi_{\vec{p}}\rangle$ is NOT an eigenstate of the momentum operator:
\begin{equation}
	\hat{\vec{p}}|\phi_{\vec{p}}\rangle=\frac{1}{\sqrt{2E_p}}\int\frac{d^{3}k}{(2\pi)^{3}}\ f(\vec{k}-\vec{p})\vec{k}|\phi(\vec{k})\rangle_{\text{QFT}}\neq\vec{p}|\phi_{\vec{p}}\rangle\ .
\end{equation}

A useful relation between the matrix element of an operator
$\hat{o}(\vec{x})$ expressed in terms of the QFT and QM states can be derived as follows. I first define its Fourier transform as:
\begin{equation}
	\hat{O}(\vec{q})\equiv\int d^{3}xe^{-i\vec{q}\cdot\vec{x}}\hat{o}(\vec{x})\ ,
\end{equation}
With this definition and the translational invariance of $\hat{o}(\vec{x})$, one can derive the following relation:
\begin{eqnarray}
	\langle\phi'_{\vec{p}_{f}}|\hat{O}(\vec{q})|\phi_{\vec{p}_{i}}\rangle & = &  \frac{1}{\sqrt{4E_iE_f}}\int\frac{d^{3}k_{i}}{(2\pi)^{3}}|f(\vec{k}_{i}-\vec{p}_{i})|^{2}\ _{\text{QFT}}\langle\phi'(\vec{k}_{i}-\vec{q})|\hat{o}(0)|\phi(\vec{k}_{i})\rangle_{\text{QFT}}\ ,
\end{eqnarray}
where $\vec{q}\equiv\vec{p}_{i}-\vec{p}_{f}$. Finally, applying Eqs.\eqref{eq:WPpeak}, \eqref{eq:WPnormalize} gives the following relation:
\begin{equation}
	\ _{\text{QFT}}\langle\phi'(\vec{p}_{f})|\hat{o}(0)|\phi(\vec{p}_{i})\rangle_{\text{QFT}}\approx\sqrt{4E_iE_f}\langle\phi'_{\vec{p}_{f}}|\hat{O}(\vec{q})|\phi_{\vec{p}_{i}}\rangle\ ,\label{eq:QFT-QM}
\end{equation}
which relates a QFT matrix element of an operator $\hat{o}(\vec{x})$ to a QM matrix element of its Fourier transform; the latter is what one can compute in nuclear theory. 

\subsection{Beta decay at tree-level }

At the scale $q\ll M_W$, the strangeness-conserving charged weak interaction Lagrangian between electrons and quarks takes the following form:
\begin{equation}
	\mathcal{L}=-\frac{G_FV_{ud}}{\sqrt{2}}\bar{e}\gamma_\mu(1-\gamma_5)\nu_e\bar{u}\gamma^\mu(1-\gamma_5)d+\text{h.c.}\ ,
\end{equation}
where $G_F$ is Fermi's constant. I now consider a nuclear $\beta^{+}$ decay: $i\rightarrow f+e^{+}+\nu_{e}$. The tree-level QFT decay amplitude takes the following form:
\begin{equation}
\mathcal{M}=-\frac{G_F V_{ud}^*}{\sqrt{2}}\bar{u}_\nu\gamma_\mu(1-\gamma_5)v_e\:{}_{\text{QFT}}\langle f(\vec{p}_f)|\bar{d}(0)\gamma^\mu(1-\gamma_5)u(0)|i(\vec{p}_i)\rangle_\text{QFT}\ .
\end{equation}
Using Eq.\eqref{eq:QFT-QM}, the QFT nuclear matrix element reduces to:
\begin{equation}
	{}_{\text{QFT}}\langle f(\vec{p}_f)|\bar{d}(0)\gamma^\mu(1-\gamma_5)u(0)|i(\vec{p}_i)\rangle_\text{QFT}=\sqrt{4E_iE_f}(H_V^\mu+H_A^\mu)\ ,
\end{equation}
where
\begin{eqnarray}
	H^{\mu}_{V} & \equiv & \int d^{3}xe^{-i\vec{q}\cdot\vec{x}}\langle f_{\vec{p}_{f}}|\bar{d}(\vec{x})\gamma^{\mu}u(\vec{x})|i_{\vec{p}_{i}}\rangle\nonumber \\
	H^{\mu}_{A} & \equiv & -\int d^{3}xe^{-i\vec{q}\cdot\vec{x}}\langle f_{\vec{p}_{f}}|\bar{d}(\vec{x})\gamma^{\mu}\gamma_{5}u(\vec{x})|i_{\vec{p}_{i}}\rangle\ .
\end{eqnarray}
The subscripts $V$ and $A$ denote the contribution from the vector and axial charged weak current, respectively. 

In allowed beta decays, the unsuppressed components of $H_{V,A}^{\mu}$ are $H^{0}_{V}$
and $\vec{H}_{A}$, which correspond to the ``Fermi'' and ``Gamow-Teller''
transition, respectively. Here I concentrate on $H^{0}_{V}$, which is the only unsuppressed term in superallowed $0^+\rightarrow 0^+$ transition. It can be expanded as:
\begin{equation}
	H^{0}_{V}  =  \int d^{3}xe^{-i\vec{q}\cdot\vec{x}}\langle f_{\vec{p}_{f}}|d^\dagger(\vec{x})u(\vec{x})|i_{\vec{p}_{i}}\rangle =  \langle f|\left(\int d^{3}xd^\dagger(\vec{x})u(\vec{x})\right)|i\rangle+\dots
\end{equation}
Here I have denoted $|\phi\rangle\equiv|\phi_{\vec{0}}\rangle$ for simplicity.
The dots in the equation above denote recoil corrections to $H^{0}_{V}$
that depend on the small nuclear momenta $\vec{p}_{i,f}$. The non-recoil piece defines the ``Fermi matrix element''
$M_{F}$:
\begin{equation}
	M_{F}\equiv\langle f|\left(\int d^{3}x q^\dagger(\vec{x})\tau_{+}q(\vec{x})\right)|i\rangle=\langle f|T_{+}|i\rangle\ ,
\end{equation}
where the corresponding ``Fermi transition operator'' $T_{+}$ is
nothing but the isospin-raising operator discussed in Sec.\ref{sec:isospin}. Therefore, any reasonable calculation of $M_F$ must be based on a construction of the operator $T_+$ that satisfies the correct isospin commutation relations. 

\section{\label{sec:T+nuclear}Constructing the Fermi transition operator at nuclear level}

Sec.\ref{sec:isospin} defines the isospin generators $\vec{T}$
in terms of quark fields. At the nuclear level, these operators are defined similarly in terms of neutron and proton field operators. Specifically, 
\begin{eqnarray}
	T_{3} & = & \frac{1}{2}\int d^{3}x\left[n^{\dagger}(\vec{x})n(\vec{x})-p^{\dagger}(\vec{x})p(\vec{x})\right]\nonumber \\
	T_{+} & = & \int d^{3}xn^{\dagger}(\vec{x})p(\vec{x})\nonumber \\
	T_{-} & = & \int d^{3}xp^{\dagger}(\vec{x})n(\vec{x})=T^{\dagger}_{+}\ ,
\end{eqnarray}
which satisfy correct isospin commutation relation $[T_{+},T_{-}]=2T_{3}$.

To express the isospin operators in a second-quantized form, one recalls the
expansion of a non-relativistic fermionic field operator $\psi(\vec{x})$ using a complete
set of single-fermion wave functions $\{\phi_{\alpha}(\vec{x})\}$:
\begin{equation}
	\psi(\vec{x})=\sum_{\alpha}\phi_{_{\alpha}}(\vec{x})a_{\alpha}\ ,
\end{equation}
where $a_{\alpha}$ is the annihilation operator of the single-fermion
state $\alpha$ (creation operator of the anti-particle is not needed for non-relativistic fields). The single-fermion wavefunctions satisfy the following orthonormality and completeness relations:
\begin{eqnarray}
	\int d^{3}x\phi^{*}_{\alpha}(\vec{x})\phi_{\beta}(\vec{x}) & = & \delta_{\alpha\beta}\nonumber \\
	\sum_{\alpha}\phi_{\alpha}(\vec{x})\phi^{*}_{\alpha}(\vec{y}) & = & \delta^{3}(\vec{x}-\vec{y})\ , \label{eq:orthocomplete}
\end{eqnarray}
while the creation and annihilation operators satisfy the following
anti-commutation relation:
\begin{equation}
	\{a_{\alpha},a^{\dagger}_{\beta}\}=\delta_{\alpha\beta}\ .
\end{equation}
This leads to the correct equal-time anti-commutation relation of
the fermionic field operator:
\begin{equation}
	\{\psi(\vec{x}),\psi^{\dagger}(\vec{y})\}=\delta^{3}(\vec{x}-\vec{y})\ .
\end{equation}
It is important to emphasize that, one has the absolute freedom to choose any set of single-fermion wavefunctions $\{\phi_\alpha(\vec{x})\}$ for the expansion, as long as they satisfy the orthonormality and completeness relation in Eq.\eqref{eq:orthocomplete}.

Following the procedure above, I expand the neutron and proton field operators as:
\begin{eqnarray}
n(\vec{x}) & = & \sum_{\alpha}\phi^{n}_{\alpha}(\vec{x})a_{n,\alpha}\nonumber \\
p(\vec{x}) & = & \sum_{\alpha}\phi^{p}_{\alpha}(\vec{x})a_{p,\alpha}\ ,\label{eq:npexpansion}
\end{eqnarray}
where $\alpha=(n_{\alpha},\ell_{\alpha},j_{\alpha},m_{\alpha})$
denotes collectively the radial and angular quantum numbers. The operators $a_{n,\alpha}$ and $a_{p,\alpha}$ annihilate a neutron and proton of the quantum number $\alpha$, respectively. The single-neutron and single-proton basis wavefunctions $\{\phi_\alpha^n(\vec{x})\}$, $\{\phi_\alpha^p(\vec{x})\}$ are separately orthonormal and complete. They can be split into the radial and angular part as: 
\begin{eqnarray}
 \phi^{n}_{\alpha}(\vec{x}) & = & R^{n}_{n_{\alpha}\ell_{\alpha}}(r)\mathcal{Y}_{\ell_{\alpha}j_{\alpha}m_{\alpha}}(\Omega)\nonumber \\
\phi^{p}_{\alpha}(\vec{x}) & = & R^{p}_{n_{\alpha}\ell_{\alpha}}(r)\mathcal{Y}_{\ell_{\alpha}j_{\alpha}m_{\alpha}}(\Omega)\ .
\end{eqnarray}
They only differ by the radial wavefunction, and I can define the ``radial overlap integral'' between the neutron and proton basis wavefunctions as:
\begin{eqnarray}
r_{\alpha\beta} & \equiv & \int d^{3}x\phi^{n*}_{\alpha}(\vec{x})\phi^{p}_{\beta}(\vec{x})\nonumber \\
& = & \int^{\infty}_{0}dr\ r^{2}R^{n}_{n_{\alpha}\ell_{\alpha}}(r)R^{p}_{n_{\beta}\ell_{\beta}}(r)\int d\Omega\mathcal{Y}^{*}_{\ell_{\alpha}j_{\alpha}m_{\alpha}}(\Omega)\mathcal{Y}_{\ell_{\beta}j_{\beta}m_{\beta}}(\Omega)\nonumber \\
& = & \delta_{\ell_{\alpha}\ell_{\beta}}\delta_{j_{\alpha}j_{\beta}}\delta_{m_{\alpha}m_{\beta}}\int^{\infty}_{0}dr\ r^{2}R^{n}_{n_{\alpha}\ell_{\alpha}}(r)R^{p}_{n_{\beta}\ell_{\alpha}}(r)\ .\label{eq:ralphabeta}
\end{eqnarray}
Once again, I emphasize that the choices of $\{\phi_\alpha^n(\vec{x})\}$ and $\{\phi_\alpha^p(\vec{x})\}$ are completely arbitrary; for instance, one can choose them to be the same, $\phi_\alpha^n(\vec{x})=\phi_\alpha^p(\vec{x})\equiv \phi_\alpha(\vec{x})$, which leads to $r_{\alpha\beta}=\delta_{\alpha\beta}$. One can also keep them distinct, in which case $r_{\alpha\alpha}\neq 1$ and $r_{\alpha\beta}\neq 0$ for $\alpha\neq\beta$.  

I now construct the isospin operators with the expansion of neutron and proton field operator in Eq.\eqref{eq:npexpansion}. The $T_{3}$ operator takes a simple form:
\begin{eqnarray}
	T_{3} & = & \frac{1}{2}\int d^{3}x\left[n^{\dagger}(\vec{x})n(\vec{x})-p^{\dagger}(\vec{x})p(\vec{x})\right]\nonumber \\
	& = & \frac{1}{2}\sum_{\alpha}\left(a^{\dagger}_{n,\alpha}a_{n,\alpha}-a^{\dagger}_{p,\alpha}a_{p,\alpha}\right)\ ,
\end{eqnarray}
which simply counts the difference between the total number of neutrons and protons. Meanwhile, the isospin raising operator takes the following form:
\begin{equation}
	T_{+} =  \int d^{3}xn^{\dagger}(\vec{x})p(\vec{x}) =  \sum_{\alpha\beta}r_{\alpha\beta}a^{\dagger}_{n,\alpha}a_{p,\beta}\ .
\end{equation}
To facilitate future discussions, I split the expression above into the $\alpha=\beta$
piece and the $\alpha\neq\beta$ piece:
\begin{eqnarray}
	T_{+} & = & \sum_{\alpha}r_{\alpha\alpha}a^{\dagger}_{n,\alpha}a_{p,\alpha}+\sum_{\alpha\neq\beta}r_{\alpha\beta}a^{\dagger}_{n,\alpha}a_{p,\beta}\nonumber \\
	& = & \sum_{\alpha}a^{\dagger}_{n,\alpha}a_{p,\alpha}-\sum_{\alpha}\Omega_{\alpha}a^{\dagger}_{n,\alpha}a_{p,\alpha}+\sum_{\alpha\neq\beta}r_{\alpha\beta}a^{\dagger}_{n,\alpha}a_{p,\beta}\label{eq:T+np}
\end{eqnarray}
where I have defined the following ``radial mismatch factor'':
\begin{equation}
	\Omega_{\alpha}\equiv1-r_{\alpha\alpha}\ .
\end{equation}
In this representation, the Fermi matrix element consists of three terms:
\begin{eqnarray}
M_F & = & \langle f|\sum_{\alpha}a^{\dagger}_{n,\alpha}a_{p,\alpha}|i\rangle-\langle f|\sum_{\alpha}\Omega_{\alpha}a^{\dagger}_{n,\alpha}a_{p,\alpha}|i\rangle+\langle f|\sum_{\alpha\neq\beta}r_{\alpha\beta}a^{\dagger}_{n,\alpha}a_{p,\beta}|i\rangle\ .\label{eq:MF3terms}
\end{eqnarray}

Now, if I take $\{\phi_\alpha^n(\vec{x})\}$ to be only slightly different from $\{\phi_\alpha^p(\vec{x})\}$, then both $\Omega_\alpha$ and $r_{\alpha\neq\beta}$ become numerically small, and the Fermi matrix element is dominated by the first term at the right hand side of Eq.\eqref{eq:MF3terms}, which is close to $\sqrt{2}$. The second term arises from the slight mismatch between the neutron and the proton radial wavefunction with the same radial quantum number, and therefore may be called the ``radial mismatch correction''. The third term arises from the fact that proton can transition into a neutron with a different radial quantum number, and therefore may be called the ``radial excitation correction''. 

\section{Examining the C1C2 formalism}

With the preparation above, one can now better understand the C1C2 formalism developed in Ref.\cite{Towner:2007np}, as well as its limitations. In the paper, the Fermi matrix element was first written as:
\begin{equation}
	M_{F}=\sum_{\alpha\beta}\langle f|a^{\dagger}_{n,\alpha}a_{p,\beta}|i\rangle\langle\alpha|\tau_{+}|\beta\rangle\ ,
\end{equation}
which is so far exact. A crucial assumption, however, was made on the single-nucleon
matrix element $\langle\alpha|\tau_{+}|\beta\rangle$:
\begin{equation}
	\langle\alpha|\tau_{+}|\beta\rangle_{\text{HT}}\equiv\delta_{\alpha\beta}r_{\alpha\alpha}\ ,
\end{equation}
where $r_{\alpha\alpha}$ is exactly the radial overlap integral I
defined in Eq.\eqref{eq:ralphabeta}, but restricted to $\alpha=\beta$. With this, their Fermi matrix element
reduces to:
\begin{eqnarray}
	(M_{F})_{\text{C1C2}} & = & \sum_{\alpha}\langle f|a^{\dagger}_{n,\alpha}a_{p,\alpha}|i\rangle r_{\alpha\alpha}\nonumber \\
	& = & \langle f|\sum_{\alpha}a^{\dagger}_{n,\alpha}a_{p,\alpha}|i\rangle-\langle f|\sum_{\alpha}\Omega_{\alpha}a^{\dagger}_{n,\alpha}a_{p,\alpha}|i\rangle\ . \label{eq:MFC1C2}
\end{eqnarray}

The two terms at the right hand side of Eq.\eqref{eq:MFC1C2} were computed in Ref.\cite{Towner:2007np} as follows:
\begin{itemize}
	\item 
 In the first term (which
is large), they took the limit $\phi_\alpha^n=\phi_\alpha^p$ (i.e. $\{a_{n,\alpha},a_{p,\alpha}\}$ transforms as an isospin doublet), but considered the fact that $|i\rangle$, $|f\rangle$
are not exact isospin eigenstates as they mix
states with different $T$. This results in a deviation of this term
from the isospin limit $M_{F}^0=\sqrt{2}$, parameterized as:
\begin{equation}
\sum_{\alpha}\langle f|a^{\dagger}_{n,\alpha}a_{p,\alpha}|i\rangle \equiv M_F^0\left(1-\frac{1}{2}\delta_{C1}\right)\ ,
\end{equation}
which defines $\delta_{C1}$ as the ``isospin mixing contribution'' to $\delta_{C}$. This term was computed in shell model with a modest model space, in which Coulomb and other charge-dependent terms are added to the Hamiltonian; the latter are adjusted case-by-case to reproduce the IMME.
\item Meanwhile, in the second term they assumed
that $|i\rangle$, $|f\rangle$ are exact isospin eigenstates, but
considered the non-zeroness of the radial mismatch factor $\Omega_{\alpha}$. This term was parameterized as:
\begin{equation}
\langle f|\sum_{\alpha}\Omega_{\alpha}a^{\dagger}_{n,\alpha}a_{p,\alpha}|i\rangle\equiv \frac{1}{2}M_F^0\delta_{C2}\ ,
\end{equation}
which defines $\delta_{C2}$ as the ``radial overlap correction'' to $\delta_{C}$. To compute this term, they adopted the proton and neutron radial wavefunctions derived from a Woods-Saxon potential, with the parameters in the potential adjusted case-by-case to reproduce the nuclear charge radius and the proton and neutron separation energies. 
\end{itemize}
Numerically, they always
found $\delta_{C2}\gg\delta_{C1}$. Combining the two terms above, the absolute square of the full Fermi matrix element reads:
\begin{equation}
	|M_F|^2_{C1C2}\approx (M_F^0)^2(1-\delta_{C1}-\delta_{C2})\ ,
\end{equation}
so in this formalism, the ISB correction $\delta_C$ is the sum of the ``isospin mixing correction'' $\delta_{C1}$ and the ``radial mismatch correction'' $\delta_{C2}$. 

Miller and Schwenk \cite{Miller:2008my} pointed out that the formalism above implies the following
form of the Fermi transition operator:
\begin{equation}
	(T_{+})_{\text{C1C2}}=\sum_{\alpha}a^{\dagger}_{n,\alpha}a_{p,\alpha}-\sum_{\alpha}\Omega_{\alpha}a^{\dagger}_{n,\alpha}a_{p,\alpha}\ ,\label{eq:T+HT}
\end{equation}
which does not satisfy the basic commutation relation $[T_+,T_-]=2T_3$.
In fact, comparing to Eq.(\ref{eq:T+np}),
one immediately see that Eq.(\ref{eq:T+HT}) misses
the radial excitation term $\sum_{\alpha\neq\beta}r_{\alpha\beta}a^{\dagger}_{n,\alpha}a_{p,\beta}$, which is needed to restore the correct commutation relation. 
If one defines the nuclear matrix element of this operator as:
\begin{equation}
\sum_{\alpha\neq\beta}r_{\alpha\beta}\langle f|a^{\dagger}_{n,\alpha}a_{p,\beta}|i\rangle\equiv -\frac{1}{2}M_F^0\delta_{C3}\ ,
\end{equation}
then, the full ISB correction $\delta_C$ should consist of three terms:
\begin{equation}
	\delta_C=\delta_{C1}+\delta_{C2}+\delta_{C3}\ .
\end{equation}
In a follow-up work~\cite{Miller:2009cg}, Miller and Schwenk
estimated the size of $\delta_{C3}$, which they found to be numerically significant and largely cancel
out $\delta_{C2}$, leading to a much smaller value of
$\delta_{C}$ (the cancellation could be even exact at the second order of the ISB potential $V$, under certain assumptions of the dominant intermediate state); this conclusion is also supported by more recent studies~\cite{Xayavong:2025api}.

Following the argument in Sec.\ref{sec:T+nuclear}, one can see that such a cancellation is no coincidence. In both the work by Towner-Hardy~\cite{Towner:2007np} and Miller-Schwenk~\cite{Miller:2008my,Miller:2009cg}, the single-nucleon wavefunctions $\{\phi_\alpha^n(\vec{x}),\phi_\alpha^p(\vec{x})\}$ are derived (to the greatest extent possible) from the ``physical'' Hamiltonian $H=H_0+V$, which leads to the scaling $\Omega_\alpha\sim V^2$, $r_{\alpha\neq\beta}\sim V$; the latter is multiplied to a transition matrix element that involves radial excitation, which picks up another factor of $V$. However, as I show in Eq.\eqref{eq:npexpansion}, these wavefunctions can be any arbitrary basis used to expand the neutron and proton field operators, which do not need to be related to $H$. In other words, the values of both $\Omega_\alpha$ and $r_{\alpha\neq \beta}$ are completely arbitrary, so they are unphysical quantities which the final value of $\delta_C$ cannot depend on. Therefore, the ``radial mismatch'' and ``radial excitation'' contribution (as well as $\delta_{C1}$) must exhibit a strong cancellation to remove all such unphysical basis-dependence, just like removing the arbitrary gauge parameter in computing QFT amplitudes in a gauge theory. The C1C2 formalism, which misses the $\delta_{C3}$ term, would upset such a cancellation, and its resulting $\delta_C$ is analogous to a gauge theory amplitude with an uncanceled gauge parameter. In fact, the observation that even with the same C1C2 + Woods-Saxon framework, different calculations can still return very different results for $\delta_{C1}$ and $\delta_{C2}$ (see, e.g. Table II in Ref.\cite{Xayavong:2025jdh}) is, to some extent, a reflection of such arbitrariness. 

\begin{figure}[tb]
	\includegraphics[scale=1]{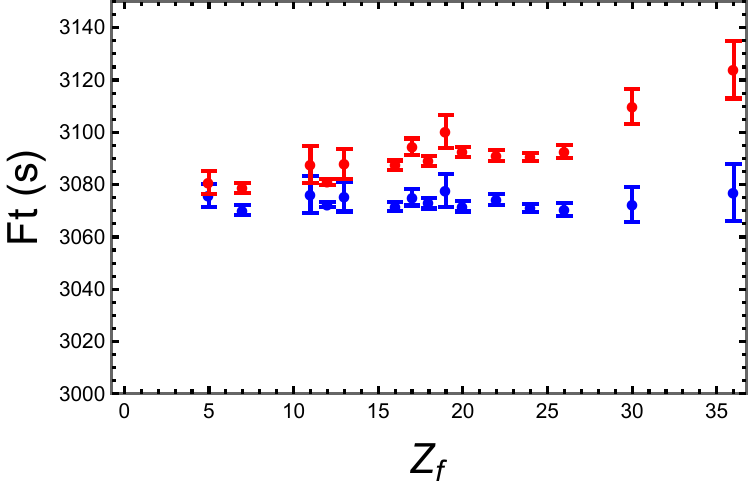}\hfill
	\caption{\label{fig:Ftalign}Strictly for illustrative purpose: the plot of the 15 most precise $\mathcal{F}t$ values taken in Ref.\cite{Hardy:2020qwl} (blue), and the same set of values with the replacement of $\delta_C\rightarrow\delta_{C1}$ in Table XIV of Ref.\cite{Hardy:2020qwl} (red). $Z_f$ indicates the charge of the daughter nucleus.}
\end{figure}

To gain more insights, one can consider a special choice of single-nucleon basis wavefunctions, which is the isospin symmetric basis $\phi_\alpha^n(\vec{x})=\phi_\alpha^p(\vec{x})\equiv\phi_\alpha(\vec{x})$. This choice can always be taken as long as it is applied simultaneously to all three terms in Eq.\eqref{eq:MF3terms}. With such a choice, the radial mismatch and radial excitation term automatically vanish, while the operators $\{a_{n,\alpha},a_{p,\alpha}\}$ in the first term now transform as an isospin doublet, so all the contributions to $\delta_C$ come from the isospin mixing of the external states $|i\rangle$, $|f\rangle$ in the first term caused by the ISB potential $V$ in the Hamiltonian.  
One immediately realizes that, this is exactly what Ref.\cite{Towner:2007np} did in computing $\delta_{C1}$. This leads to a natural speculation that, the magnitude of the true value of $\delta_C$ should be closer to $\delta_{C1}$ in the C1C2 formalism, instead of $\delta_{C1}+\delta_{C2}$ which is much larger. 
To have a flavor of what it could imply, I show in Fig.\ref{fig:Ftalign} a plot of 15 best $\mathcal{F}t$ values in Ref.\cite{Hardy:2020qwl}, in comparison to the same set of values but with $\delta_C\rightarrow\delta_{C1}$. One observes that, apart from the two heaviest transitions (which can have large systematic uncertainties in $\delta_{C1}$), the latter tends to flatten out at medium mass and curve downward with decreasing $Z_f$, which can be interpreted as a signature of a BSM-induced scalar current. While the plot is strictly for illustrative purpose, its main message is that one should keep an open mind to the possibility that a more refined calculation of $\delta_C$ could lead to a misalignment of $\mathcal{F}t$ values. If the latter occurs, it should not be taken as a defect, but rather a new opportunity in probing BSM physics.  

\section{Perturbative formalism of $\delta_C$}

After identifying the fundamental limitations in the present state-of-the-art determination of $\delta_C$, it is essential to develop new computational strategies to overcome such shortcomings. It is important that the new strategy is executable using cutting-edge nuclear \textit{ab initio} methods, in order to ensure a controlled theoretical uncertainty of the outputs.

As I argued in the introduction, extracting $\delta_C$ directly from the full Fermi matrix element essentially implies a precision goal of $0.01\%$ for nuclear many-body calculations of $M_F$ with the full Hamiltonian $H=H_0+V$, which is practically impossible either due to statistical uncertainties (for stochastic methods) or various truncations that are unavoidable in different \textit{ab initio} methods but can introduce spurious ISB effects at the same level or even larger than the physical $\delta_C$. The perturbative approach offers a satisfactory solution to all the challenges above, as it directly tackles the small, physical $\delta_C$ instead of the big $M_F$, which avoids subtracting two big numbers to obtain a small number. Assuming $\delta_C\sim 10^{-3}$, this would mean in the perturbative approach one needs only a $10\%$ theory precision, instead of $0.01\%$, to achieve our final precision goal for the $V_{ud}$ extraction. In addition, all the intrinsic inaccuracies inherited from the many body methods --- statistical uncertainties or spurious ISB effects --- are automatically pushed to higher order by construction, in contrast to the brute force calculation of $M_F$ during which such inaccuracies can be at the same order and inseparable from the physical ISB signal. 

In this section, I start with a brief recap of the theory foundations introduced in Refs.\cite{Seng:2022epj,Seng:2023cvt}, and proceed by relaxing some of the assumptions made in those works on $V$. This gives rise to a more general perturbative expression of $\delta_C$.

\subsection{The $\mathcal{O}(V^{2})$ expansion}

I start with the famous Wigner-Brillouin expression, an exact relation between the full state $|n\rangle$ and the unperturbed state $|n)$:
\begin{equation}
	|n\rangle =\sqrt{\mathcal{Z}_n}[1+(E_n-\Lambda_n H\Lambda_n)^{-1}\Lambda_n V]|n)\ ,
\end{equation} 
where readers can find in Appendix \ref{sec:WBpert} the definition of various notations and the derivation of this expression. Applying the formula above to the definition of the full Fermi matrix element, $M_F=\langle f|T_+|i\rangle$ gives:
\begin{equation}
	M_{F}  =   M^{0}_{F}\sqrt{\mathcal{Z}_{i}\mathcal{Z}_{f}}\left[1+\frac{1}{M^{0}_{F}}(f|V\Lambda_{f}\frac{1}{E_{f}-\Lambda_{f}H\Lambda_{f}}T_{+}\frac{1}{E_{i}-\Lambda_{i}H\Lambda_{i}}\Lambda_{i}V|i)\right]\ .
	\label{eq:MFexact}
\end{equation}
Taking its absolute square and comparing to Eq.\eqref{eq:deltaCdef} gives the exact expression of $\delta_C$ as follows:
\begin{equation}
	\delta_{C}=1-\mathcal{Z}_{i}\mathcal{Z}_{f}\left|1+\frac{1}{M^{0}_{F}}(f|V\Lambda_{f}\frac{1}{E_{f}-\Lambda_{f}H\Lambda_{f}}T_{+}\frac{1}{E_{i}-\Lambda_{i}H\Lambda_{i}}\Lambda_{i}V|i)\right|^{2}\ ,
\end{equation}
where $M_F^0=\sqrt{2}$. Expanding the expression above to the leading non-vanishing terms in powers of $V$ gives~\cite{Seng:2023cvt}:
\begin{eqnarray}
\delta_{\text{C}} & \approx & (i|V\Lambda_{g}\frac{1}{(E^{0}_{g}-\Lambda_{g}H_{0}\Lambda_{g})^{2}}\Lambda_{g}V|i)+(f|V\Lambda_{g}\frac{1}{(E^{0}_{g}-\Lambda_{g}H_{0}\Lambda_{g})^{2}}\Lambda_{g}V|f)\nonumber \\
&  & -\sqrt{2}\mathfrak{Re}(f|V\Lambda_{g}\frac{1}{E^{0}_{g}-\Lambda_{g}H_{0}\Lambda_{g}}T_{+}\frac{1}{E^{0}_{g}-\Lambda_{g}H_{0}\Lambda_{g}}\Lambda_{g}V|i)\ ,
\end{eqnarray}
which is at the order $\mathcal{O}(V^2)$, in accordance to the Behrends-Sirlin-Ademollo-Gatto theorem~\cite{Behrends:1960nf,Ademollo:1964sr}. Here I have modified the definition of the projection operator in the perturbative expression to 
\begin{equation}
\Lambda_{g}\equiv1-\sum^{+1}_{T_{3}=-1}|g;1,T_{3})(g;1,T_{3}|
\end{equation}
to project away the entire $|g;1)$ isotriplet, just to make the operator isospin-invariant.
Finally, to facilitate the later discussion of isospin decomposition, I move $T_+$ in the second line across the isospin-invariant operator $(E_g^0-\Lambda_g H_0 \Lambda_g)^{-1}\Lambda_g$ and commute it with $V$. Recalling that $T_+|i)=\sqrt{2}|f)$, the outcome reads:
\begin{eqnarray}
\delta_{\text{C}} & \approx & (i|V\Lambda_{g}\frac{1}{(E^{0}_{g}-\Lambda_{g}H_{0}\Lambda_{g})^{2}}\Lambda_{g}V|i)-(f|V\Lambda_{g}\frac{1}{(E^{0}_{g}-\Lambda_{g}H_{0}\Lambda_{g})^{2}}\Lambda_{g}V|f)\nonumber \\
&  & -\sqrt{2}\mathfrak{Re}(f|V\Lambda_{g}\frac{1}{(E^{0}_{g}-\Lambda_{g}H_{0}\Lambda_{g})^2}\Lambda_{g}[T_+,V]|i)\ .\label{eq:deltaCpert}
\end{eqnarray}
This is the starting point of the perturbative calculation of $\delta_{C}$. One can imagine inserting a complete set of eigenstates $\{|X)\}$ into each term, which converts the operator $(E_g^0-\Lambda_gH_0\Lambda_g)^{-2}$ into $(E_g^0-E_X^0)^{-2}$. Since there are two energy denominators, the contribution to $\delta_C$ from excited states with $E_X^0\gg E_g^0$ will be heavily suppressed, so one expects Eq.\eqref{eq:deltaCpert} to be saturated by the contribution from the low-lying $J^P=0^+$ excited states (because $V$ conserves angular momentum and parity); specific studies, e.g. Ref.\cite{Auerbach:2008ut}, suggest that the isovector monopole states play an important role.

\subsection{Isospin decomposition of $V$}

The presence of $T_+$ in the second line in Eq.\eqref{eq:deltaCpert} complicates the problem, because its commutator with $V$ does not take a simple form without specifying the isospin content of the latter. The ISB potential $V$ can be decomposed into spherical tensors $V^T_{T_3}$ in the isospin space:
\begin{equation}
V=V_0^1+V_0^2+\dots\ .
\end{equation}
For $n$-body (nucleon) operators, the expansion truncates at $T=n$. In Refs.\cite{Seng:2022epj,Seng:2023cvt}, only the $V_0^1$ term was retained, and was assumed to take the form of a  Coulomb potential from a uniformly-charged sphere. In this work I will relax such approximations to make the result more general. 

From the IMME (which will be further discussed later), the hierarchy $V_0^1\gg V_0^2\gg \dots$ is observed, and the truncation at $T=2$ is sufficient to describe the nuclear mass splitting. One of the main reasons is that the primary contributor of ISB effects, namely the Coulomb repulsion between protons, is a two-body operator that contributes up to $T=2$. In fact, $V_0^{T=3}$ only exists if three- or higher-body operators are included. These operators themselves are non-Coulombic and are chirally suppressed comparing to one- and two-body operators, and the $T=3$ ISB effects on top of them are even further suppressed. In addition, in most low-energy nuclear physics applications, three-nucleon forces are truncated at normal-ordered two-body level, which eliminates all the $T\geq 3$ components~\cite{Rothman:2025rua}. Therefore, for all practical purposes, it is sufficient to assume the truncation of the ISB potential at $T=2$, which I will adopt in this work:
\begin{equation}
	H=H_0+V_0^1+V_0^2\ .\label{eq:Htruncate}
\end{equation}
Extension to higher $T$ is straightforward, but would not be necessary unless there are strong numerical indications that these higher-order terms can contribute at the level of $10^{-4}$ or more. 

The isolation of individual $V_0^T$ from the full Hamiltonian $H$ can be done using the following SU(2) identity in the isospin space~\cite{Sakurai:2011zz}:
\begin{equation}
[T_{\pm},V^{T}_{T_{3}}]=\sqrt{(T\mp T_{3})(T\pm T_{3}+1)}V^{T}_{T_{3}\pm1}\ .\label{eq:SU2commutation}
\end{equation}
With the identity above and the assumed ISB truncation at $T=2$ in Eq.\eqref{eq:Htruncate}, one can derive the following relations:
\begin{eqnarray}
V^{1}_{0} & = & \frac{1}{2}[T_{+},[T_{-},H]]-\frac{1}{8}[T_{+},[T_{+},[T_{-},[T_{-},H]]]]\nonumber \\
V^{2}_{0} & = & \frac{1}{24}[T_{+},[T_{+},[T_{-},[T_{-},H]]]]\ ,
\end{eqnarray}
which express the ISB potentials in terms of commutators between the full Hamiltonian and the isospin ladder operators that can be performed numerically. In addition to the $T_3=0$ component, one is interested also in the $T_3=-1$ component of $V_{T_3}^T$ as will be seen later. They can be obtained from $V_0^T$ using Eq.\eqref{eq:SU2commutation}, or directly from the Hamiltonian (assuming the truncation in Eq.\eqref{eq:Htruncate}) as:
\begin{eqnarray}
V^{1}_{-1} & = & \frac{1}{\sqrt{2}}\left\{ [T_{-},H]-\frac{1}{4}[T_{+},[T_{-},[T_{-},H]]]\right\} \nonumber \\
V^{2}_{-1} & = & \frac{1}{4\sqrt{6}}[T_{+},[T_{-},[T_{-},H]]]\ .\label{eq:Vm1}
\end{eqnarray}
Also, notice that $V_{+1}^T=-(V_{-1}^T)^\dagger$.

Among all the contributors to ISB, the Coulomb potential between point-like protons in a nucleus is an important component which is worth special attention. It can be written as:
\begin{equation}
V_C=\frac{1}{2}\sum_{i\neq j}\frac{\alpha}{r_{ij}}\left(\frac{1}{2}-\frac{1}{2}\tau_{z}(i)\right)\left(\frac{1}{2}-\frac{1}{2}\tau_{z}(j)\right)=\frac{\alpha}{8}\sum_{i\neq j}\frac{1}{r_{ij}}\left(1-\tau_{z}(i)\right)\left(1-\tau_{z}(j)\right)\ ,
\end{equation}
where $r_{ij}\equiv|\vec{r}_{i}-\vec{r}_{j}|$ and $\alpha$ is the
fine structure constant. One can split this potential into $T=0,1,2$
components, $V_C=(V_C)^{0}_{0}+(V_C)^{1}_{0}+(V_C)^{2}_{0}$, where
\begin{eqnarray}
(V_C)^{0}_{0} & = & \frac{\alpha}{8}\sum_{i\neq j}\frac{1}{r_{ij}}\left(1+\frac{1}{3}\vec{\tau}(i)\cdot\vec{\tau}(j)\right)\nonumber \\
(V_C)^{1}_{0} & = & -\frac{\alpha}{8}\sum_{i\neq j}\frac{1}{r_{ij}}\left(\tau_{z}(i)+\tau_{z}(j)\right)\nonumber \\
(V_C)^{2}_{0} & = & \frac{\alpha}{8}\sum_{i\neq j}\frac{1}{r_{ij}}\left(\tau_{z}(i)\tau_{z}(j)-\frac{1}{3}\vec{\tau}(i)\cdot\vec{\tau}(j)\right)\ .
\end{eqnarray}
Meanwhile, the $T_{3}=-1$ components read:
\begin{eqnarray}
(V_C)^{1}_{-1} & = & -\frac{\sqrt{2}\alpha}{8}\sum_{i\neq j}\frac{1}{r_{ij}}\left(\tau_{-}(i)+\tau_{-}(j)\right)\nonumber \\
(V_C)^{2}_{-1} & = & \frac{\sqrt{6}\alpha}{24}\sum_{i\neq j}\frac{1}{r_{ij}}\left(\tau_{-}(i)\tau_{z}(j)+\tau_{z}(i)\tau_{-}(j)\right)\ .
\end{eqnarray}
Since these potentials are analytically known, they serve as useful benchmarks in comparing the results of perturbative $\delta_C$ calculations with different \textit{ab initio} approaches.

Another important remark is as follows. In perturbation theory, the ground-state matrix elements of the ISB potential give rise to the nuclear mass splitting (see Sec.\ref{sec:IMME} for more discussions). The electromagnetically-induced nuclear mass shift is obtained from the self-energy diagram, which takes the following form:
\begin{equation}
	\int\frac{d^4q}{(2\pi)^4}\frac{T^\mu_{\:\:\mu}(p,q)}{q^2+i\epsilon}\ ,
\end{equation}
where $T^{\mu\nu}(p,q)$ is the electromagnetic Compton tensor that involves a time-ordered product of two electromagnetic currents. It turns out that the amplitude of neutrinoless double beta decay ($0\nu\beta\beta$) has exactly the same integral structure, except that the electromagnetic currents are replaced by the weak currents. It was shown that, in an effective field theory description of the $0\nu\beta\beta$, the nuclear matrix element of the long-distance $\Delta T_3=-2$ potential has an ultraviolet divergence, which must be absorbed by a contact potential that encodes short-distance physics~\cite{Cirigliano:2018hja}. The low energy constants (LECs) of this contact term is unknown, which constitutes a large theory uncertainty in the prediction of the $0\nu\beta\beta$ rate. Given the similarity of the integral structure, it is apparent that similar counterterms must so exist in $V$. Below I write down the corresponding short-distance potentials in the $T=1$ and $T=2$ channel:
\begin{eqnarray}
(V_\text{sd})_0^1&=&a_1\sum_{i\neq j}\delta^3(\vec{r}_i-\vec{r}_j)(\tau_z(i)+\tau_z(j))\nonumber\\
(V_\text{sd})_0^2&=&a_2\sum_{i\neq j}\delta^3(\vec{r}_i-\vec{r}_j)\left(\tau_z(i)\tau_z(j)-\frac{1}{3}\vec{\tau}(i)\cdot\vec{\tau}(j)\right)\ ,\label{eq:Vsd}
\end{eqnarray}
and their $T_3=-1$ component:
\begin{eqnarray}
	(V_\text{sd})_{-1}^1&=&\sqrt{2}a_1\sum_{i\neq j}\delta^3(\vec{r}_i-\vec{r}_j)(\tau_{-}(i)+\tau_{-}(j))\nonumber\\
	(V_\text{sd})_{-1}^2&=&\frac{\sqrt{6}}{3}a_2\sum_{i\neq j}\delta^3(\vec{r}_i-\vec{r}_j)\left(\tau_-(i)\tau_z(j)+\tau_z(i)\tau_-(j)\right)\ .
\end{eqnarray}
The two nucleus-independent LECs $\{a_1,a_2\}$ can be fitted to ISB observables in experiment and will be discussed in Sec.\ref{sec:IMME}.

\subsection{Isospin analysis of $\delta_{C}$\label{sec:deltaCisospin}}

The truncation of ISB potential at $T=2$ implies that the $\mathcal{O}(V^{2})$ expression of
$\delta_{C}$ splits into three terms:
\begin{equation}
	\delta_{C}=\delta^{(1,1)}_{C}+\delta^{(1,2)}_{C}+\delta^{(2,2)}_{C}\ ,\label{eq:deltacsplit}
\end{equation}
where $\delta^{(1,1)}_{C}=\mathcal{O}((V^{1}_{0})^{2})$, $\delta^{(1,2)}_{C}=\mathcal{O}(V^{1}_{0}V^{2}_{0})$,
$\delta^{(2,2)}_{C}=\mathcal{O}((V^{2}_{0})^{2})$ . Due to the hierarchy of the ISB potentials, one expects
$|\delta^{(1,1)}_{C}|>|\delta^{(1,2)}_{C}|>|\delta^{(2,2)}_{C}|$. In this subsection, I analyze each of these terms using isospin symmetry, which eventually recast them into a compact form in terms of reduced matrix elements in the isospin space.

First, the operators involved in Eq.\eqref{eq:deltaCpert} can be seen as products of two operators: $V^T_{T_3}\Lambda_g$ and $(E_g^0-\Lambda_g H_0\Lambda_g)^{-2}\Lambda_g V^{T'}_{T_3'}$ which are rank $T$ and  $T'$ tensors in the isospin space, respectively. I  can express their product as:
\begin{equation}
V^T_{T_3}\Lambda_g\frac{1}{(E_g^0-\Lambda_g H_0 \Lambda_g)^2}\Lambda_g V^{T'}_{T_3'}=\sum_{T'',T_3''}C_{T,T_3;T',T_3'}^{T'',T_3''}\Gamma_{T_3''}^{T,T';T''}\ ,\label{eq:Gammatensor}
\end{equation}
where $C_{T,T_3;T',T_3'}^{T'',T_3''}$ are Clebsch-Gordan (CG) coefficients, and $\Gamma_{T_3''}^{T,T';T''}$ is a rank $T''$ tensor in the isospin space. Using Wigner-Eckart theorem in isospin space, I can express its matrix element with respect to the $|g;1,T_3)$ states in terms of reduced matrix elements:
\begin{equation}
(g;1,T_{3f}|\Gamma_{T_3''}^{T,T';T''}|g;1,T_{3i})=C_{1,T_{3i};T'',T_3''}^{1,T_{3f}}(g;1||\Gamma^{T,T';T''}||g;1)\ .\label{eq:reduced}
\end{equation}
A useful relation of the reduced matrix element with respect to $T\leftrightarrow T'$ reads:
\begin{equation}
(g;1||\Gamma^{T,T';T''}||g;1)=(g;1||\Gamma^{T',T;T''}||g;1)^*\ ,\label{eq:conjugate}
\end{equation}
which proof can be found in Appendix \ref{sec:symmetryproof}.

Using the above and the fact that $T_{3f}=T_{3i}+1$ and $T_{3i}=-1,0$, I can write the different components of $\delta_C$ in terms of the reduced matrix elements. The results read:
\begin{eqnarray}
\delta_C^{(1,1)}&=&(i|V_0^1\Lambda_{g}\frac{1}{(E^{0}_{g}-\Lambda_{g}H_{0}\Lambda_{g})^{2}}\Lambda_{g}V_0^1|i)-(f|V_0^1\Lambda_{g}\frac{1}{(E^{0}_{g}-\Lambda_{g}H_{0}\Lambda_{g})^{2}}\Lambda_{g}V_0^1|f)\nonumber \\
&  & -\mathfrak{Re}(f|V_0^1\Lambda_{g}\frac{2}{(E^{0}_{g}-\Lambda_{g}H_{0}\Lambda_{g})^2}\Lambda_{g}V_{+1}^1|i)\nonumber\\
&=&-(g;1||\Gamma^{1,1;1}||g;1)\ . \label{eq:delta11}
\end{eqnarray}
\begin{eqnarray}
	\delta_{\text{C}}^{(1,2)} & = & (i|V_0^1\Lambda_{g}\frac{1}{(E^{0}_{g}-\Lambda_{g}H_{0}\Lambda_{g})^{2}}\Lambda_{g}V_0^2|i)+(i|V_0^2\Lambda_{g}\frac{1}{(E^{0}_{g}-\Lambda_{g}H_{0}\Lambda_{g})^{2}}\Lambda_{g}V_0^1|i)\nonumber\\
	&&-(f|V_0^1\Lambda_{g}\frac{1}{(E^{0}_{g}-\Lambda_{g}H_{0}\Lambda_{g})^{2}}\Lambda_{g}V_0^2|f)-(f|V_0^2\Lambda_{g}\frac{1}{(E^{0}_{g}-\Lambda_{g}H_{0}\Lambda_{g})^{2}}\Lambda_{g}V_0^1|f)\nonumber \\
	&  & -\mathfrak{Re}(f|V_0^1\Lambda_{g}\frac{\sqrt{12}}{(E^{0}_{g}-\Lambda_{g}H_{0}\Lambda_{g})^2}\Lambda_{g}V_{+1}^2|i)-\mathfrak{Re}(f|V_0^2\Lambda_{g}\frac{2}{(E^{0}_{g}-\Lambda_{g}H_{0}\Lambda_{g})^2}\Lambda_{g}V_{+1}^1|i)\nonumber\\
	&=&-2\sqrt{\frac{3}{5}}(\delta_{T_{3i},0}-\delta_{T_{3i},-1})\mathfrak{Re}(g;1||\Gamma^{1,2;2}||g;1)\ .\label{eq:delta12}
\end{eqnarray}
\begin{eqnarray}
\delta_C^{(2,2)}&=&(i|V_0^2\Lambda_{g}\frac{1}{(E^{0}_{g}-\Lambda_{g}H_{0}\Lambda_{g})^{2}}\Lambda_{g}V_0^2|i)-(f|V_0^2\Lambda_{g}\frac{1}{(E^{0}_{g}-\Lambda_{g}H_{0}\Lambda_{g})^{2}}\Lambda_{g}V_0^2|f)\nonumber \\
&  & -\mathfrak{Re}(f|V_0^2\Lambda_{g}\frac{\sqrt{12}}{(E^{0}_{g}-\Lambda_{g}H_{0}\Lambda_{g})^2}\Lambda_{g}V_{+1}^2|i)\nonumber\\
&=&\frac{3}{\sqrt{5}}(g;1||\Gamma^{2,2;1}||g;1)\ .\label{eq:delta22}
\end{eqnarray}
One observes that $\delta_C^{(1,1)}$ and $\delta_{C}^{(2,2)}$ are independent of $T_{3i}$, but $\delta_{C}^{(1,2)}$ depends on it. The conclusion in Refs.\cite{Seng:2022epj,Seng:2023cvt} that $\delta_C$ is constant across the same isotriplet is only true in the $V_0^{2}\rightarrow 0$ limit.


\section{Generating functions of $\delta_{C}$}

The main task in the perturbative formalism is to evaluate the reduced matrix elements in Eqs.\eqref{eq:delta11}-\eqref{eq:delta22}. They arise from a second-order perturbation theory and thus involve two energy denominators, i.e. two factors of the nuclear Green's functions $G(z)\equiv 1/(z-H)$. In this section I define
a series of ``generating functions'' of the form:
\begin{equation}
F(z)\sim \langle \phi|V G(z) V|\phi\rangle\ ,
\end{equation} 
where $z$ is a free energy variable. This form which involves only a single Green's
function is familiar to the \textit{ab initio} community, for example in the study of nuclear response functions through the Lorentz Integral transform~\cite{Efros:1994iq,Efros:2007nq}, and more recently, the nucleus-dependent radiative corrections in beta decays~\cite{Gennari:2024sbn}. Subsequently, $\delta_{C}$ can be obtained as the first derivative
of these generating functions with respect to $z$. 

From the analysis in Sec.\ref{sec:deltaCisospin}, the involved tensor operators in $\delta_C$ are $\Gamma^{1,1;1}$, $\Gamma^{1,2;2}$, and $\Gamma^{2,2;1}$. Given that $C_{1,0;1,0}^{1,0}=C_{1,0;2,0}^{2,0}=C_{2,0;2,0}^{1,0}=0$, products of the form $V_0^TG(z)V_0^{T'}$ will never give rise to the needed tensor operators. Therefore, I construct the generating function with operator products of the form $(V_{-1}^T)^\dagger G(z) V_{-1}^{T'}$ (which is why $V_{-1}^T$ was discussed above).

\subsection{$\delta^{(1,1)}_{C}$}

To evaluate $\delta_C^{(1,1)}$, I define the following generating function:
\begin{eqnarray}
	F^{(1)}_{T_{3}}(z)&\equiv&\langle (g;T_{3})^{\perp,1}|G(z)|(g;T_{3})^{\perp,1}\rangle\nonumber\\
	|(g;T_{3})^{\perp,1}\rangle&\equiv& V_{-1}^1|g;T_3\rangle-|g;T_3-1\rangle \langle g;T_3-1|V_{-1}^1|g;T_3\rangle\ .
\end{eqnarray}
The second term in $|(g;T_{3})^{\perp,1}\rangle$ serves to remove the $0^+$ ground state after the action of $V_{-1}^1$. This removal is necessary only for $T_{3}=+1,0$ but not for $T_{3}=-1$. In the above, both $G(z)$ and the states are defined with respect to the full Hamiltonian $H$; this is just for practical simplicity, one could also choose to define them with respect to the isospin-symmetric Hamiltonian $H_0$, which impact on $\delta_C$ is of higher order in $V$. Taking the derivative of the generating function and retaining only the $\mathcal{O}(V^2)$ piece gives (remember $(V_{-1}^T)^\dagger=-V_{+1}^T$):
\begin{eqnarray}
\left.\frac{d}{dz}F^{(1)}_{T_{3}}(z)\right|_{z=E_{g,T_3}} & \approx & (g;1,T_{3}|V^{1}_{+1}\Lambda_g\frac{1}{(E_g^0-\Lambda_gH_{0}\Lambda_g)^2}\Lambda_gV^{1}_{-1}|g;1,T_{3})\nonumber \\
&=&\frac{1}{\sqrt{3}}C_{1,T_3;0,0}^{1,T_3}(g;1||\Gamma^{1,1;0}||g;1)+\frac{1}{\sqrt{2}}C_{1,T_3;1,0}^{1,T_3}(g;1||\Gamma^{1,1;1}||g;1)\nonumber\\
&&+\frac{1}{\sqrt{6}}C_{1,T_3;2,0}^{1,T_3}(g;1||\Gamma^{1,1;2}||g;1)\ .
\end{eqnarray}
This leads to:
\begin{equation}
	\delta^{(1,1)}_{C}\approx\left.\frac{d}{dz}F^{(1)}_{-1}(z)\right|_{z=E_{g;-1}}-\left.\frac{d}{dz}F^{(1)}_{+1}(z)\right|_{z=E_{g;+1}}\ .\label{eq:master1}
\end{equation} 

\subsection{$\delta^{(2,2)}_{C}$}

To evaluate $\delta_C^{(2,2)}$, I define the following generating function:
\begin{eqnarray}
	F^{(2)}_{T_{3}}(z)&\equiv&\langle (g;T_{3})^{\perp,2}|G(z)|(g;T_{3})^{\perp,2}\rangle\nonumber\\
	|(g;T_{3})^{\perp,2}\rangle&\equiv& V_{-1}^2|g;T_3\rangle-|g;T_3-1\rangle \langle g;T_3-1|V_{-1}^2|g;T_3\rangle\ .
\end{eqnarray}
Taking its derivative and retaining only the $\mathcal{O}(V^2)$ piece gives:
\begin{eqnarray}
\left.\frac{d}{dz}F^{(2)}_{T_{3}}(z)\right|_{z=E_{g,T_3}} & \approx & (g;1,T_{3}|V^{2}_{+1}\Lambda_g\frac{1}{(E_g^0-\Lambda_gH_{0}\Lambda_g)^2}\Lambda_gV^{2}_{-1}|g;1,T_{3})\nonumber \\
&=&-\frac{1}{\sqrt{5}}C_{1,T_3;0,0}^{1,T_3}(g;1||\Gamma^{2,2;0}||g;1)-\frac{1}{\sqrt{10}}C_{1,T_3;1,0}^{1,T_3}(g;1||\Gamma^{2,2;1}||g;1)\nonumber\\
&&+\frac{1}{\sqrt{14}}C_{1,T_3;2,0}^{1,T_3}(g;1||\Gamma^{2,2;2}||g;1)\ .
\end{eqnarray}
This leads to:
\begin{equation}
\delta^{(2,2)}_{C}\approx3\left.\frac{d}{dz}F^{(2)}_{-1}(z)\right|_{z=E_{g;-1}}-3\left.\frac{d}{dz}F^{(2)}_{+1}(z)\right|_{z=E_{g;+1}}\ .\label{eq:master2}
\end{equation}

\subsection{$\delta^{(1,2)}_{C}$}

To evaluate $\delta_{C}^{(1,2)}$, I define the following generating function:
\begin{eqnarray}
	F^{(x)}_{T_{3}}(z)&\equiv&\langle (g;T_{3})^{\perp,1+2}|G(z)|(g;T_{3})^{\perp,1+2}\rangle-F_{T_3}^{(1)}(z)-F_{T_3}^{(2)}(z)\nonumber\\
	|(g;T_{3})^{\perp,1+2}\rangle&\equiv& (V_{-1}^1+V_{-1}^2)|g;T_3\rangle-|g;T_3-1\rangle \langle g;T_3-1|(V_{-1}^1+V_{-1}^2)|g;T_3\rangle\ .
\end{eqnarray}
Taking its derivative and retaining only the $\mathcal{O}(V^2)$ piece gives:
\begin{eqnarray}
\left.\frac{d}{dz}F^{(x)}_{T_{3}}(z)\right|_{z=E_{g,T_3}} & \approx & (g;1,T_{3}|V^{1}_{+1}\Lambda_g\frac{1}{(E_g^0-\Lambda_gH_{0}\Lambda_g)^2}\Lambda_gV^{2}_{-1}|g;1,T_{3})\nonumber \\
&&+(g;1,T_{3}|V^{2}_{+1}\Lambda_g\frac{1}{(E_g^0-\Lambda_gH_{0}\Lambda_g)^2}\Lambda_gV^{1}_{-1}|g;1,T_{3})\nonumber\\
&=&2\sqrt{\frac{3}{10}}C_{1,T_3;1,0}^{1,T_3}\mathfrak{Re}(g;1||\Gamma^{1,2;1}||g;1)+\sqrt{2}C_{1,T_3;2,0}^{1,T_3}\mathfrak{Re}(g;1||\Gamma^{1,2;2}||g;1)\ .\nonumber\\
\end{eqnarray}
This leads to:
\begin{eqnarray}
	\delta^{(1,2)}_{C}  &\approx & -\sqrt{3}(\delta_{T_{3i},0}-\delta_{T_{3i},-1})\left\{ \left.\frac{d}{dz}F^{(x)}_{-1}(z)\right|_{z=E_{g;-1}}+\left.\frac{d}{dz}F^{(x)}_{+1}(z)\right|_{z=E_{g;+1}}\right\} \nonumber \\
	&\approx &\sqrt{3}(\delta_{T_{3i},0}-\delta_{T_{3i},-1}) \left.\frac{d}{dz}F^{(x)}_{0}(z)\right|_{z=E_{g;0}} \ .\label{eq:masterx}
\end{eqnarray}

With the above, the perturbation calculation of $\delta_C$ simply proceeds as follows: one first computes the generating functions using standard techniques such as the Lanczos
algorithm~\cite{Lanczos:1950zz,Haydock_1974,Marchisio:2002jx}, then takes their first derivative and sets $z$ to the energy of the $0^+$ state associated with each generating function. Finally, substituting the results into Eqs.\eqref{eq:master1}, \eqref{eq:master2}, \eqref{eq:masterx} returns the quantities  $\delta_C^{(T,T')}$, which combine to give the full $\delta_C$ through Eq.\eqref{eq:deltacsplit}.

\section{\label{sec:IMME}Benchmarking $V$ with IMME coefficients}

The accuracy of the perturbative calculation of $\delta_C$ formulated above relies on the successful modeling of two ingredients in the generating function: (1) the ISB potential $V$, and (2) the nuclear Green's function $G(z)$. In the following two sections, I outline the methods to benchmark the \textit{ab initio} treatment of these two ingredients with experimental observables. 

The accuracy of the $V$-modeling can be benchmarked with experimental results of the IMME, which encodes the ISB effects to nuclear masses:
\begin{equation}
M(T_{3})=a+bT_{3}+cT^{2}_{3}\ ,\label{eq:IMME}
\end{equation}
where $M(T_{3})$ is the mass of the nucleus in an isospin multiplet
with the third isospin quantum number $T_{3}$. The coefficients $b$
and $c$ quantify the ISB effects and are well measured in experiments~\cite{Lam:2013bhc}. Using the Hamiltonian in Eq.\eqref{eq:Htruncate}, the nuclear mass splitting for the ground-state $0^+$ isotriplet can be computed from first-order perturbation theory as:
\begin{eqnarray}
\Delta M(T_{3}) & = &
 (g;1,T_{3}|V^{1}_{0}+V^{2}_{0}|g;1,T_{3})\nonumber \\
& = & C^{1,T_{3}}_{1,T_{3};1,0}(g;1||V^{1}||g;1)+C^{1,T_{3}}_{1,T_{3};2,0}(g;1||V^{2}||g;1)\ ,\label{eq:IMMEpert}
\end{eqnarray}
where the relevant CG coefficients are:
\begin{equation}
C^{1,T_{3}}_{1,T_{3};1,0}=   \frac{T_{3}}{\sqrt{2}}\ ,\ 
C^{1,T_{3}}_{1,T_{3};2,0}  = \frac{3T^{2}_{3}-2}{\sqrt{10}}\ .
\end{equation}
Comparing Eq.\eqref{eq:IMMEpert} to Eq.\eqref{eq:IMME} gives the coefficients $b$ and $c$ in terms of the ground-state reduced matrix elements of $V^T_{T_3}$:
\begin{equation}
b  =  \frac{(g;1||V^{1}||g;1)}{\sqrt{2}}\ ,\ 
c  =  \frac{3(g;1||V^{2}||g;1)}{\sqrt{10}}\ .\label{eq:bccoef}
\end{equation}
The fact that in experiment one always observes $|b|\gg |c|$ indicates the hierarchy $V_0^1\gg V_0^2$. 

Eq.\eqref{eq:bccoef} allows the benchmarking of $V$ using experimental results of $b$, $c$ by computing their $0^+$ ground state matrix element. Specifically, 
\begin{equation}
b  =  \langle g;1|V^{1}_{0}|g;1\rangle=  -\langle g;-1|V^{1}_{0}|g;-1\rangle =  \langle g;0|V^{1}_{-1}|g;1\rangle =  \langle g;-1|V^{1}_{-1}|g;0\rangle
\end{equation}
and 
\begin{align}
c  &=  3\langle g;1|V^{2}_{0}|g;1\rangle =  -\frac{3}{2}\langle g;0|V^{2}_{0}|g;0\rangle= 3\langle g;-1|V^{2}_{0}|g;-1\rangle =  \sqrt{3}\langle g;0|V^{2}_{-1}|g;1\rangle\nonumber\\
&=  -\sqrt{3}\langle g;-1|V^{2}_{-1}|g;0\rangle = \frac{3}{\sqrt{6}}\langle g;-1|V^{2}_{-2}|g;1\rangle\ .
\end{align}
In particular, comparing the ground state matrix element of $V_{-1}^T$ with the IMME coefficients are useful as $V_{-1}^T$ are the components of the ISB potential that actually enter the generating functions.

Below are some remarks on the short-distance potential $V_\text{sd}$ defined in Eq.\eqref{eq:Vsd}. The two LECs $\{a_1,a_2\}$ must be determined before the perturbative calculation of $\delta_C$ can be performed reliably. This can be done by tuning their values to reproduce the IMME coefficients $b$, $c$ in a given nuclear isotriplet. Since the LECs are nucleus-independent, a consistency check is then to use these determined $\{a_1,a_2\}$ to predict the IMME coefficients in other nuclear isotriplets, and compare with the experimental results. In additions, these LECs can also be related to ISB combinations of two-nucleon scattering lengths $a_{NN'}$; specifically, $a_1$ can be inferred from $a_{pp}-a_{nn}$, and $a_2$ from $a_{pp}+a_{nn}-2a_{pn}$. In fact, the $T=2$ case was already studied in Ref.\cite{Cirigliano:2018hja,Cirigliano:2020dmx} for $0\nu\beta\beta$, where $a_2$ here is proportional to $C_1+C_2$ in that paper.

\section{Benchmarking $V$ and $G(z)$ with ISB in nuclear charge radii}

Benchmarking the \textit{ab initio} results of ground state matrix elements of $V$ against $b$, $c$ alone is not sufficient to ensure the reliability of the outcome of $\delta_C$ computed with the same method, because the latter depends also on the nuclear Green's function $G(z)$ that connects the ground state to $0^+$ excited states through $V$. Therefore, one needs to identify additional experimental observable that allows for a simultaneous benchmarking of $V$ and $G(z)$, on top of the benchmarking procedure with the IMME. 

In this section, I will develop further from the proposal in Ref.\cite{Seng:2022epj}, which considered the following combination of the mean-square (MS) charge radius of the $0^+$ ground state isotriplet:
\begin{equation}
	\Delta M^{(1)}_{B}\equiv\frac{1}{2}\left(Z_{+1}\langle r^{2}_{\text{ch}}\rangle_{+1}+Z_{-1}\langle r^{2}_{\text{ch}}\rangle_{-1}\right)-Z_{0}\langle r^{2}_{\text{ch}}\rangle_{0}\ ,\label{eq:DeltaMB1}
\end{equation}
where $\langle r_\text{ch}^2\rangle_{T_3}$ is the MS charge radius of the state $|g;T_3\rangle$, which is a measurable quantity. One can show that $\Delta M_B^{(1)}$ vanishes in the $V\rightarrow 0$ limit, so it probes ISB effects in nuclear charge radii. In what follows, I will derive an expression of $\Delta M_B^{(1)}$ in first-order perturbation theory, which demonstrates its simultaneous dependence on $V$ and $G(z)$. 

To do so, I start with the point-like nucleon approximation, where the MS
charge radius is obtained as:
\begin{equation}
	Z_{T_3}\langle r^{2}_{\text{ch}}\rangle_{T_3}=\sum^{A}_{i=1}\langle g;T_3|\frac{1}{2}r^{2}_{i}\left(1-\tau_z(i)\right)|g;T_3\rangle\equiv\langle g;T_3|(\mathbb{R}^{2})^{0}_{0}+(\mathbb{R}^{2})^{1}_{0}|g;T_3\rangle\ .\label{eq:rchleading}
\end{equation}
The isoscalar and isovector MS radii operators are defined as:
\begin{equation}
	(\mathbb{R}^{2})^{0}_{0}\equiv\frac{1}{2}\sum^{A}_{i=1}r^{2}_{i}\ ,\ (\mathbb{R}^{2})^{1}_{0}\equiv-\frac{1}{2}\sum^{A}_{i=1}r^{2}_{i}\tau_z(i)\ .
\end{equation}
In reality, there are corrections to the expression above from finite
nucleon size, Darwin-Foldy term, spin-orbit term, meson-exchange current,
and so on. But the ISB corrections on top of these already small terms will be doubly-suppressed, so as far as $\Delta M_B^{(1)}$ is concerned, it is sufficient to take Eq.\eqref{eq:rchleading} as the starting point. 

The ISB effect in Eq.\eqref{eq:rchleading} resides in the state $|g;T_3\rangle$ which, to $\mathcal{O}(V)$, is related to the isospin eigenstate $|g;1,T_3)$ as (see Appendix \ref{sec:WBpert}):
\begin{equation}
	|g;T_3\rangle\approx\left(1+\frac{1}{E^{0}_{g}-\Lambda_{g}H_{0}\Lambda_{g}}\Lambda_g V\right)|g;1,T_3)\ .
\end{equation}
With this, Eq.\eqref{eq:rchleading} can be expanded to $\mathcal{O}(V)$:
\begin{eqnarray}
	Z_{T_{3}}\langle r^{2}_{\text{ch}}\rangle_{T_{3}} 
	& \approx & \sum_{T=0,1}\Biggl[(g;1,T_{3}|(\mathbb{R}^{2})^{T}_{0}|g;1,T_{3})\nonumber \\
	&  & \left.+\sum_{T'=1,2}\left\{ (g;1,T_{3}|(\mathbb{R}^{2})^{T}_{0}\frac{1}{E^{0}_{g}-\Lambda_{g}H_{0}\Lambda_{g}}\Lambda_g V^{T'}_{0}|g;1,T_{3})+\text{c.c.}\right\} \right]\ .\nonumber \\ \label{eq:rch2pert}
\end{eqnarray}
Given that $(\mathbb{R}^{2})^{T}_{0}$ is a rank $T$ tensor and $(E^{0}_{g}-\Lambda_{}H_{0}\Lambda_{g})^{-1}\Lambda_gV^{T'}_{0}$
is a rank $T'$ tensor in isospin space, I can express their product as:
\begin{equation}
	(\mathbb{R}^{2})^{T}_{0}\frac{1}{E^{0}_{g}-\Lambda_{g}H_{0}\Lambda_{g}}\Lambda_gV^{T'}_{0}=\sum_{T''}C^{T'',0}_{T,0;T',0}\tilde{\Gamma}^{T,T';T''}_{0}
\end{equation}
where $\tilde{\Gamma}^{T,T';T''}_{0}$ is a rank-$T''$ tensor. With this, I can rewrite Eq.\eqref{eq:rch2pert} as:
\begin{eqnarray}
	Z_{T_{3}}\langle r^{2}_{\text{ch}}\rangle_{T_{3}} & \approx & \sum_{T=0,1}\Biggl[(g;1,T_{3}|(\mathbb{R}^{2})^{T}_{0}|g;1,T_{3})\nonumber \\
	&  & \left.+\sum_{T'=1,2}\left\{ \sum_{T''}C^{T'',0}_{T,0;T',0}(g;1,T_{3}|\tilde{\Gamma}^{T,T';T''}_{0}|g;1,T_{3})+\text{c.c.}\right\} \right]\nonumber \\
	& = & \sum_{T=0,1}\Biggl[C^{1,T_{3}}_{1T_{3};T,0}(g;1||(\mathbb{R}^{2})^{T}||g;1)\nonumber \\
	&  & \left.+\sum_{T'=1,2}\left\{ \sum_{T''}C^{1,T_{3}}_{1,T_{3};T'',0}C^{T'',0}_{T,0;T',0}(g;1||\tilde{\Gamma}^{T,T';T''}||g;1)+\text{c.c.}\right\} \right]\ ,\nonumber \\
\end{eqnarray}
where in the second equation I have used the Wigner-Eckart theorem to define the reduced matrix elements $(g;1||(\mathbb{R}^2)^T||g;1)$ and $(g;1||\tilde{\Gamma}^{T,T';T''}||g;1)$, the former is independent of $V$ and the latter is linear to $V$. Evaluating this equation for $T_{3}=-1,0,+1$ and substituting the result to Eq.\eqref{eq:DeltaMB1} gives:
\begin{equation}
	\Delta M^{(1)}_{B}\approx\sqrt{\frac{3}{5}}(g;1||\tilde{\Gamma}^{1,1;2}||g;1)+\frac{3}{\sqrt{10}}(g;1||\tilde{\Gamma}^{0,2;2}||g;1)+\text{c.c.}\ ,\label{eq:DeltaMB1pert}
\end{equation}
where the $V$-independent terms completely cancel out as expected, and only two reduced matrix elements at $\mathcal{O}(V)$ survive, the first induced
by $V^{1}_{0}$ and the second induced by $V^{2}_{0}$. 

The task now is to construct diagonal matrix elements that give rise to the reduced matrix elements in Eq.\eqref{eq:DeltaMB1pert}. For the $V^{1}_{0}$ contribution, I start with:
\begin{equation}
	(\mathbb{R}^{2})^{1}_{0}\frac{1}{E^{0}_{g}-\Lambda_{g}H_{0}\Lambda_{g}}\Lambda_gV^{1}_{0}=-\frac{1}{\sqrt{3}}\tilde{\Gamma}^{1,1;0}_{0}+\sqrt{\frac{2}{3}}\tilde{\Gamma}^{1,1;2}_{0}\ .
\end{equation}
So, $(g;1||\tilde{\Gamma}^{1,1;2}||g;1)$ can be obtained by combining two diagonal matrix elements of the equation above. An example is:
\begin{eqnarray}
	\sqrt{\frac{3}{5}}(g;1||\tilde{\Gamma}^{1,1;2}||g;1)+\text{c.c.} & = & (g;1,1|(\mathbb{R}^{2})^{1}_{0}\frac{1}{E^{0}_{g}-\Lambda_{g}H_{0}\Lambda_{g}}\Lambda_gV^{1}_{0}|g;1,1)\nonumber \\
	&  & -(g;1,0|(\mathbb{R}^{2})^{1}_{0}\frac{1}{E^{0}_{g}-\Lambda_{g}H_{0}\Lambda_{g}}\Lambda_gV^{1}_{0}|g;1,0)+\text{c.c.}\nonumber \\ \label{eq:r2V01}
\end{eqnarray}
Next, for the $V^{2}_{0}$ contribution, I start with:
\begin{equation}
	(\mathbb{R}^{2})^{0}_{0}\frac{1}{E^{0}_{g}-\Lambda_{g}H_{0}\Lambda_{g}}\Lambda_gV^{2}_{0}=\tilde{\Gamma}^{0,2;2}_{0}\ .
\end{equation}
So, $(g;1||\tilde{\Gamma}^{0,2;2}||g;1)$ can be obtained by taking any diagonal matrix element of the equation above. An example is:
\begin{eqnarray}
	\frac{3}{\sqrt{10}}(g;1||\tilde{\Gamma}^{0,2;2}||g;1)+\text{c.c.} & = & 3(g;1,1|(\mathbb{R}^{2})^{0}_{0}\frac{1}{E^{0}_{g}-\Lambda_{g}H_{0}\Lambda_{g}}\Lambda_gV^{2}_{0}|g;1,1)+\text{c.c.}\nonumber \\ \label{eq:r2V02}
\end{eqnarray}

Eqs.\eqref{eq:r2V01} and \eqref{eq:r2V02}, which are the starting point for a perturbative calculation of $\Delta M_B^{(1)}$, involve matrix elements that are asymmetric in left and right operators, in analogy to $\delta_C^{(1,2)}$. It can be symmetrized using the polarization identity $A\cdot B+B\cdot A=(A+B)\cdot (A+B)-A\cdot A-B\cdot B$, but $(\mathbb{R}^{2})^{T}$
and $V^{T'}_{0}$ cannot be directly added because they have different
dimensions. To add them, I can multiply $(\mathbb{R}^{2})^{T}$ by a
dimensionful constant factor, preferably a constant to make its size
comparable to $V^{T}_{0}$. For that purpose, I recall the ``uniformly
charged sphere'' approximation to the nuclear charge distribution,
which results in the following Coulomb potential within the sphere~\cite{Seng:2022epj}:
\begin{equation}
V_{\text{sphere}}=-\frac{Z\alpha}{4R^{3}_{C}}\sum^{A}_{i=1}\left(r^{2}_{i}-3R^{2}_{C}\right)\left(1-\tau_z(i)\right)\ ,
\end{equation}
where $R_{C}=\sqrt{5/3}\times1.1\text{fm}\times A^{1/3}$. Inspired
by this form, I define the operator:
\begin{equation}
\mathbb{V}^{\mathbb{T}}_{0}\equiv\frac{Z\alpha}{4R^{3}_{C}}(\mathbb{R}^{2})^{\mathbb{T}}_{0}\ 
\end{equation}
which now has the same dimension as $V^{T}_{0}$. While $Z$ and $A$
(hidden in $R_{C}$) can be freely chosen, the most natural choice
is of course just the $Z$ and $A$ of the external nuclear state.

With these, I can now define the following symmetric matrix elements, in analogy to the generating functions for $\delta_C$, except that $\Delta M_B^{(1)}$ arises from a first-order perturbation theory so no derivative is needed.
\begin{eqnarray}
	\mathfrak{F}_{1}(\mathbb{T};T_{3}) & \equiv & \langle (g;T_{3})^{\perp,\mathbb{T}}|G(E_{g;T_3})|(g;T_{3})^{\perp,\mathbb{T}}\rangle\nonumber\\
	|(g;T_{3})^{\perp,\mathbb{T}}\rangle&\equiv& \mathbb{V}_0^\mathbb{T}|g;T_3\rangle-|g;T_3\rangle \langle g;T_3|\mathbb{V}_0^\mathbb{T}|g;T_3\rangle
\end{eqnarray}
\begin{eqnarray}
	\mathfrak{F}_{2}(T';T_{3}) & \equiv & \langle (g;T_{3})^{\perp,T'}|G(E_{g;T_3})|(g;T_{3})^{\perp,T'}\rangle\nonumber\\
	|(g;T_{3})^{\perp,T'}\rangle&\equiv& V_0^{T'}|g;T_3\rangle-|g;T_3\rangle \langle g;T_3|V_0^{T'}|g;T_3\rangle
\end{eqnarray}
\begin{eqnarray}
	\mathfrak{F}_{x}(\mathbb{T},T';T_{3}) & \equiv & \langle (g;T_{3})^{\perp,\mathbb{T}T'}|G(E_{g;T_3})|(g;T_{3})^{\perp,\mathbb{T}T'}\rangle-\mathfrak{F}_1(\mathbb{T};T_3)-\mathfrak{F}_2(T',T_3)\nonumber\\
	|(g;T_{3})^{\perp,\mathbb{T}T'}\rangle&\equiv& (\mathbb{V}_0^{\mathbb{T}}+V_0^{T'})|g;T_3\rangle-|g;T_3\rangle \langle g;T_3|(\mathbb{V}_0^{\mathbb{T}}+V_0^{T'})|g;T_3\rangle\ .
\end{eqnarray}
With the above, the desired asymmetric matrix elements can be constructed as:
\begin{equation}
\mathfrak{M}(T,T';T_{3})\equiv\frac{4R^{3}_{C}}{Z\alpha}\mathfrak{F}_{\text{x}}(\mathbb{T},T';T_{3})\ .
\end{equation}
The matrix element $\mathfrak{M}$ is defined with respect to the full Hamiltonian for simplicity; to the order $\mathcal{O}(V)$, it gives:
\begin{equation}
\mathfrak{M}(\mathbb{T},T';T_{3})\approx(g;1,T_{3}|(\mathbb{R}^{2})^{\mathbb{T}}_{0}\frac{1}{E^{0}_{g}-\Lambda_{g}H_{0}\Lambda_{g}}\Lambda_gV^{T'}_{0}|g;1,T_3)+\text{c.c.}
\end{equation}
The above and Eqs.\eqref{eq:r2V01}, \eqref{eq:r2V02} combine to provide the leading perturbative expression of $\Delta M^{(1)}_{B}$
as:
\begin{equation}
\Delta M^{(1)}_{B}\approx \mathfrak{M}(1,1;1)-\mathfrak{M}(1,1;0)+3\mathfrak{M}(0,2;1)\ .\label{eq:DeltaMmaster}
\end{equation}
The first two terms at the right hand side is the $V^{1}_{0}$ contribution,
and the third term is the $V^{2}_{0}$ contribution. 

Eq.\eqref{eq:DeltaMmaster} is the master formula for the perturbative calculation of the ISB effect in nuclear MS charge radii. The matrix element $\mathfrak{M}(\mathbb{T},T';T_3)$ can be calculated with the same \textit{ab initio} method one uses to compute $\delta_C$; since it depends not only on $V$ but also on the nuclear Green's function, it provides a benchmark for the reliability of the \textit{ab initio} treatment of $G(z)$ which is not available from the IMME. 

The feasibility of this benchmarking process relies on the experimental knowledge of $\Delta M_B^{(1)}$, which can be determined only if all the three charge radii of the $J^p=0^+$ isotriplet are measured. The ability to contribute to the benchmarking of $\delta_C$ provides an important motivation for future experiments of nuclear charge radii.
Currently, among all the nuclear isotriplets with measured superallowed transitions, only the $A=38$ isotriplet has all the three charge radii measured~\cite{Seng:2023cgl}, with $\Delta M_B^{(1)}=-0.03(54)\text{ fm}^2$. A particularly interesting isotriplet is $A=26$, where a large ISB signal in the charge radii is predicted from mirror nuclei analysis~\cite{Ohayon:2024dwt}. Therefore, the determination of the only missing charge radius, namely that of ${}^{26}\text{Si}$, is of high interest. This can be accomplished through an isotope shift measurement through laser spectroscopy at, for example, the Facility for Rare Isotope Beams (FRIB).

\section{Conclusion}

This work lays out in detail a comprehensive theory framework to compute the ISB correction $\delta_C$ to the Fermi matrix element of superallowed nuclear beta decays, a central theory input for the determination of $V_{ud}$. First, I show that the commonly-adopted values of $\delta_C$ from the $\delta_C=\delta_{C1}+\delta_{C2}$ formalism fails to satisfy the nucleon-basis-independence requirement and therefore cannot escape from large systematic errors; the large ``radial mismatch'' correction $\delta_{C2}$, in particular, is unphysical and its effect has to be canceled by the missing ``radial excitation'' correction to restore the basis-independence. Furthermore, I argue that a potential misalignment of $\mathcal{F}t$ should not be immediately interpreted as a drawback of a particular $\delta_C$ calculation, but rather an opportunity to probe new physics.

Obtaining $\delta_C$ to the required precision level from a direct calculation of the full Fermi matrix element $M_F$ implies a theory precision goal of 0.01\% for the latter, which is practically impossible. This is the largest motivation for adopting the perturbative formalism: by tackling directly the leading contributions to $\delta_C$ at $\mathcal{O}(V^2)$, one completely evades the need of subtracting two big numbers to obtain a small number, and thus reduces the precision goal for the nuclear many-body calculation tremendously to $\sim 10\%$. Through isospin analysis, I define a number of ``generating functions'' $F(z)\sim \langle \phi|VG(z)V|\phi\rangle$, which form is familiar to the nuclear \textit{ab initio} community, and their first derivatives give the various isospin components of $\delta_C$.   

The accuracy of the perturbative calculation of $\delta_C$ relies on the correct handling of the ISB potential $V$ and the nuclear Green's function $G(z)$ in the \textit{ab initio} calculation. In this paper I outline the procedure to benchmark them with two classes of experimental observables: (1) The IMME coefficients which benchmark $V$, and (2) The ISB in nuclear MS charge radii which benchmark $G(z)$ (and $V$). For each of such experimental observables, I derive their corresponding expression in the leading perturbative expansion which should be computed with the same \textit{ab initio} method used for $\delta_C$. Experimental measurements of $J^p=0^+$ isotriplet nuclear charge radii are of high interest due to their unique roles in the $\delta_C$ benchmarking process.

This theory paper represents the first step towards a thorough re-evaluation of $\delta_C$ with the perturbative approach. A follow-up work on the actual numerical calculation with multiple \textit{ab initio} methods (including NCSM and IMSRG) is currently in progress.

\begin{acknowledgments}
	
The author thank Michael Gennari, Mikhail Gorchtein and Bingcheng He for many inspiring discussions. This work is supported in part by the U.S. Department of Energy (DOE) Topical Collaboration ``Nuclear Theory for New
Physics'', award No. DE-SC0023663, and by University of Tennessee, Knoxville.

\end{acknowledgments}

\begin{appendix}

\section{\label{sec:WBpert}The Wigner-Brillouin perturbation theory}

In this appendix I provide a proof of the Wigner-Brillouin perturbation theory that connects the full state $|n\rangle$, which is an eigenstate of the full Hamiltonian $H=H_0+V$, to the unperturbed state $|n)$ which is an eigenstate of the unperturbed Hamiltonian $H_0$. To do so, I first write
\begin{equation}
	|n\rangle=\sqrt{\mathcal{Z}_{n}}\left[|n)+\sum_{m\neq n}a_{mn}|m)\right]\label{eq:nstate}
\end{equation}
where $a_{mn}$ are constant coefficients, given by $\sqrt{\mathcal{Z}_{n}}a_{mn}=(m|n\rangle$
for $m\neq n$. Now, since $|n\rangle$ is an eigenstate of $H$, it satisfies:
\begin{equation}
	E_{n}|n\rangle=H|n\rangle=\sqrt{\mathcal{Z}_{n}}\left[H|n)+\sum_{m\neq n}a_{mn}H|m)\right]\ .
\end{equation}
For any $k\neq n$, I obtain:
\begin{eqnarray}
	\sqrt{\mathcal{Z}_{n}}E_{n}a_{kn} & = & E_{n}(k|n\rangle=\sqrt{\mathcal{Z}_{n}}\left[(k|H|n)+\sum_{m\neq n}a_{mn}(k|H|m)\right]\nonumber \\
	\implies E_{n}a_{kn} & = & (k|H|n)+\sum_{m\neq n}a_{mn}(k|H|m)\ .\label{eq:Enakn}
\end{eqnarray}
The first term at the right hand side of Eq.(\ref{eq:Enakn}) can
be simplified as:
\begin{equation}
	(k|H|n)=(k|H_{0}+V|n)=E^{0}_{n}(k|n)+(k|V|n)=(k|V|n)
\end{equation}
since $k\neq n$. 

Next, I introduce the projection operator $\Lambda_{n}\equiv1-|n)(n|$
which annihilates $|n)$ but leaves all other unperturbed eigenstates
untouched:
\begin{equation}
	\Lambda_{n}|n)=0\ ,\ \Lambda_{n}|k)=|k)\ \forall\ k\neq n\ .
\end{equation}
It obviously satisfies $\Lambda^{2}_{n}=\Lambda_{n}$, and $[\Lambda_{n},\Lambda_{n}A\Lambda_{n}]=0$
for any operator $A$. With this, one can write $(k|V|n)=(k|\Lambda_{n}V|n)$
(for $k\neq n$), and the second term at the right hand side of Eq.(\ref{eq:Enakn})
can be written as:
\begin{equation}
	\sum_{m\neq n}a_{mn}(k|H|m)=\sum_{m}a_{mn}(k|\Lambda_{n}H\Lambda_{n}|m)
\end{equation}
where the summation of $m$ is no longer restricted after introducing
the projection operator. Therefore, Eq.(\ref{eq:Enakn}) becomes:
\begin{equation}
	\sum_{m}(k|E_{n}-\Lambda_{n}H\Lambda_{n}|m)a_{mn}=(k|\Lambda_{n}V|n)\ ,
\end{equation}
which can be viewed as a matrix equation:
\begin{equation}
	(E_{n}-\Lambda_{n}H\Lambda_{n})\cdot a=\Lambda_{n}V\ .
\end{equation}
So I can invert the matrix to obtain the constant coefficients:
\begin{equation}
	a_{mn}=(m|(E_{n}-\Lambda_{n}H\Lambda_{n})^{-1}\Lambda_{n}V|n)\ .
\end{equation}
Plugging this back into Eq.(\ref{eq:nstate}) gives: 
\begin{eqnarray}
	|n\rangle & = & \sqrt{\mathcal{Z}_{n}}\left[|n)+\sum_{m\neq n}|m)(m|(E_{n}-\Lambda_{n}H\Lambda_{n})^{-1}\Lambda_{n}V|n)\right]\nonumber \\
	& = & \sqrt{\mathcal{Z}_{n}}\left[|n)+\sum_{m}|m)(m|\Lambda_{n}(E_{n}-\Lambda_{n}H\Lambda_{n})^{-1}\Lambda_{n}V|n)\right]\nonumber \\
	& = & \sqrt{\mathcal{Z}_{n}}\left[1+\Lambda_{n}(E_{n}-\Lambda_{n}H\Lambda_{n})^{-1}\Lambda_{n}V\right]|n)\nonumber \\
	& = & \sqrt{\mathcal{Z}_{n}}\left[1+(E_{n}-\Lambda_{n}H\Lambda_{n})^{-1}\Lambda_{n}V\right]|n)\ , \label{eq:PT}
\end{eqnarray}
which is the desired equation. In the second last line I have used
the completeness relation $\sum_{m}|m)(m|=1$, while in the last line
I have used the fact that $\Lambda_{n}$ commutes with $(E_{n}-\Lambda_{n}H\Lambda_{n})^{-1}$
since it commutes with $E_{n}-\Lambda_{n}H\Lambda_{n}$.
Finally, the value of $\mathcal{Z}_{n}$ is simply obtained by normalization:
\begin{eqnarray}
	1 & = & \langle n|n\rangle\nonumber \\
	& = & \mathcal{Z}_{n}\left[(n|+(n|\left[(E_{n}-\Lambda_{n}H\Lambda_{n})^{-1}\Lambda_{n}V\right]^{\dagger}\right]\left[|n)+(E_{n}-\Lambda_{n}H\Lambda_{n})^{-1}\Lambda_{n}V|n)\right]\nonumber \\
	& = & \mathcal{Z}_{n}\left[(n|+(n|V\Lambda_{n}(E_{n}-\Lambda_{n}H\Lambda_{n})^{-1}\right]\left[|n)+(E_{n}-\Lambda_{n}H\Lambda_{n})^{-1}\Lambda_{n}V|n)\right]\nonumber \\
	& = & \mathcal{Z}_{n}\left[1+(n|V\Lambda_{n}(E_{n}-\Lambda_{n}H\Lambda_{n})^{-2}\Lambda_{n}V|n)\right]\nonumber \\
	\implies\mathcal{Z}_{n} & = & \left[1+(n|V\Lambda_{n}\left(E_{n}-\Lambda_{n}H\Lambda_{n}\right)^{-2}\Lambda_{n}V|n)\right]^{-1}\ .
\end{eqnarray}
Note that it expands as $\mathcal{Z}_{n}=1+\mathcal{O}(V^{2})$.

\section{\label{sec:symmetryproof}Symmetry relation between the reduced matrix elements}

Here I prove the symmetry relation $(g;1||\Gamma^{T,T';T''}||g;1)=(g;1||\Gamma^{T',T;T''}||g;1)^*$ for the reduced matrix elements defined in Eq.\eqref{eq:reduced}. I start with the identity:
\begin{eqnarray}
	&&(g;1,T_3|(V^T_{-1})^\dagger\Lambda_g\frac{1}{(E_g^0-\Lambda_g H_0 \Lambda_g)^2}\Lambda_g V^{T'}_{-1}|g;1,T_3)\nonumber\\
	&=&(g;1,T_3|(V^{T'}_{-1})^\dagger\Lambda_g\frac{1}{(E_g^0-\Lambda_g H_0 \Lambda_g)^2}\Lambda_g V^{T}_{-1}|g;1,T_3)^*\ .
\end{eqnarray}
Applying Eq.\eqref{eq:Gammatensor} to both sides yields:
\begin{equation}
	\sum_{T''}C_{T,1;T',-1}^{T'',0}(g;1,T_3|\Gamma^{T,T';T''}_0|g;1,T_3)=\sum_{T''}C_{T',1;T,-1}^{T'',0}(g;1,T_3|\Gamma^{T',T;T''}_0|g;1,T_3)^*\ .\label{eq:symmetrystep}
\end{equation}
Using the symmetry relations of the CG-coefficients:
\begin{equation}
	C_{j1,m1;j2,m2}^{j,m}=(-1)^{j1+j2-j}C_{j1,-m1;j2,-m2}^{j,-m}=(-1)^{j1+j2-j}C_{j2,m2;j1,m1}^{j,m}\ ,
\end{equation}
I obtain:
\begin{equation}
	C_{T',1;T,-1}^{T'',0}=(-1)^{T+T'-T''}C_{T,-1;T',1}^{T'',0}=C_{T,1;T',-1}^{T'',0}\ .
\end{equation}
Therefore the CG-coefficients at both sides of Eq.\eqref{eq:symmetrystep} are the same. In addition, the equality should hold for each $T''$, which implies:
\begin{equation}
	(g;1,T_3|\Gamma^{T,T';T''}_0|g;1,T_3)=(g;1,T_3|\Gamma^{T',T;T''}_0|g;1,T_3)^*\ .
\end{equation}
Finally, I express both sides in terms of reduced matrix elements using Eq.\eqref{eq:reduced}. The resulting CG-coefficients are again the same, which implies:
\begin{equation}
	(g;1||\Gamma^{T,T';T''}||g;1)=(g;1||\Gamma^{T',T;T''}||g;1)^*\ .
\end{equation}

\end{appendix}

\bibliography{ref}

\end{document}